\documentclass[aps,prd,twocolumn,superscriptaddress,showpacs,showkeys,preprintnumbers]{revtex4}

\usepackage[dvips]{graphicx}
\usepackage{epsbox,color}
\usepackage{amsmath}
\usepackage{float}
\def\slash#1{\not\!#1}

\begin{document}

\preprint{YITP-09-68}
\preprint{RBRC-822}

\title{Possible quantum numbers of the pentaquark $\Theta^+(1540)$ in QCD sum rules}


\author{Philipp Gubler}
\email{phil@th.phys.titech.ac.jp}
\affiliation{Department of Physics, H-27, Tokyo Institute of Technology, Meguro, Tokyo 152-8551, Japan}

\author{Daisuke Jido}
\affiliation{Yukawa Institute for Theoretical Physics, Kyoto University, Kyoto 606-8502, Japan}

\author{Toru Kojo}
\affiliation{RBRC, Brookhaven National Laboratory, Upton, NY 11973-5000, USA}

\author{Tetsuo Nishikawa}
\affiliation{Faculty of Health Science, Ryotokuji University, Urayasu, Chiba, 279-8567, Japan}

\author{Makoto Oka}
\affiliation{Department of Physics, H-27, Tokyo Institute of Technology, Meguro, Tokyo 152-8551, Japan}

\date{\today}

\begin{abstract}
The QCD sum rule technique is employed to investigate pentaquark states with strangeness $S = +1$ and 
$IJ^{\pi} = 0\frac{1}{2}^{\pm},1\frac{1}{2}^{\pm},0\frac{3}{2}^{\pm},1\frac{3}{2}^{\pm}$. Throughout the calculation, 
emphasis is laid on the establishment of a valid Borel window, which corresponds to a region of the Borel mass, where 
the operator product expansion converges and the presumed ground state pole dominates the sum rules. Such a Borel 
window is achieved by constructing the sum rules from the difference of two independent correlators and by 
calculating the operator product expansion up to dimension 14. 
Furthermore, we discuss the possibility of the contamination of the sum 
rules by possible $KN$ scattering states. As a result, we conclude that the $0\frac{3}{2}^{+}$ state seems to be the 
most probable candidate for the experimentally observed $\Theta^{+}(1540)$, while we also obtain states with
$0\frac{1}{2}^{-},1\frac{1}{2}^{-},1\frac{3}{2}^{+}$ at somewhat higher mass regions. 

\end{abstract}


\maketitle

\section{Introduction}
$\Theta^+(1540)$ with strangeness $S = +1$ and baryon number $B = +1$ is evidently a flavor exotic state with 
minimal quark content $uudd\overline{s}$ \cite{Diakonov}. The first announcement of its experimental detection was made in 
2003 \cite{Nakano1}, and it has since confronted the hadron physics community with interesting novel phenomena 
and unanticipated problems that have not been solved until the present day.

Presently, the experimental situation of $\Theta^+(1540)$ seems to be rather unclear. After the CLAS collaboration has published 
several papers on their pentaquark search with high statistics \cite{Battaglieri,McKinnon,Niccolai,DeVita}, where no 
signal of $\Theta^+$ could be found, many people now seem to believe that the pentaquark does not exist after all 
and that the whole story was just ``a curious episode in the history of science" \cite{Amsler}. There are, however 
still experiments that claim to observe a signal of $\Theta^+(1540)$ \cite{Barmin2,Nakano2} and therefore this issue 
should not be considered to be completely settled yet. Additional experimental results, which either unambiguously 
confirm the existence of $\Theta^+(1540)$ or otherwise can eliminate it completely, are eagerly waited for.

Theoretically, one not yet well understood property of $\Theta^+(1540)$ is its unnaturally narrow width, 
which was reported to be even less than $1\,\mathrm{MeV}$ \cite{Barmin2} and which 
is very difficult to explain from our experience 
with ordinary baryons. Because $\Theta^+(1540)$ lies about $100\,\mathrm{MeV}$ above the $KN$ threshold, one would expect 
the width to be much larger than the experimentally measured value. 
Of course, there have been many attempts to explain this narrow width of 
$\Theta^+(1540)$ \cite{Jaffe,Hosaka,Bicudo,Kishimoto,Llanes,Capstick,Karliner,Takeuchi}, but none of these approaches 
has completely succeeded yet.

Another problem is the correct assignment of quantum numbers such as spin and parity to the $\Theta^+(1540)$ state. 
There are many studies, in which states with various quantum numbers were investigated using QCD sum rules 
\cite{Zhu,Matheus1,Sugiyama,Eidemuller,Ioffe,Kondo,Nishikawa2,Matheus2,Lee,Kojo1,Gubler} or lattice QCD 
\cite{Sasaki,Ishii1,Takahashi,Lasscock1,Ishii2,Lasscock2}, but no consistent understanding has yet emerged. 

The main subject of the present paper is to determine the quantum numbers (isospin, spin, and parity) of $\Theta^+(1540)$ from 
a QCD sum rule approach. 
We therefore study and compare the sum rules of states with $IJ^{\pi} = 0\frac{1}{2}^{\pm},1\frac{1}{2}^{\pm},0\frac{3}{2}^{\pm},1\frac{3}{2}^{\pm}$.
From this comparison, we aim to determine which of the 
investigated quantum numbers is the one that most likely has to be assigned to the $\Theta^+(1540)$ state. Furthermore, 
we also look for possible excited states below $2\,\mathrm{GeV}$, that may be found in future experiments. For 
these purposes we use an improved version of the QCD sum rule method, which has first been proposed in 
\cite{Kojo1}. The basic idea of improvement is to use the difference of two correlators to construct the sum rule. 
The continuum part of the spectral function is 
significantly suppressed by this procedure, which therefore helps to find a valid Borel 
window, whose existence is a necessary condition for obtaining reliable results within the QCD sum rule technique. 
Moreover, we calculate the operator product expansion (OPE) up to dimension 14, which is indispensable for a sufficient convergence of the 
expansion.

The paper is organized as follows. In Sec. II, the formalism of QCD sum rules is briefly reviewed. 
The details of our method, including the interpolating 
fields employed and the implications of the improvement mentioned in the last 
paragraph are then explained in Sec. III. In Sec. IV, the results of the analysis for the various quantum numbers are given in detail. 
These results are then discussed in Sec. V and the conclusion is given in Sec. VI. 
Finally, the appendix is devoted to the 
numerical results of the OPE and to the details of the establishment of the Borel window for the various sum rules. 

\section{Formalism}
\subsection{QCD sum rules}
In the QCD sum rule method \cite{Shifman,Reinders}, we compute the two-point function of various operators. 
It is defined as 
\begin{equation}
\begin{split}
\Pi(q) &= i \int d^{4}x e^{iqx} \langle0|T[\eta(x) \overline{\eta}(0)]|0\rangle \\
&\equiv \Pi_1(q^2) \slash{q} + \Pi_2(q^2), \label{eq:cor1}
\end{split}
\end{equation} 
where $\eta(x)$ is a spin $\frac{1}{2}$ operator. 
$\Pi_1(q^2)$ is called the chiral-even, and $\Pi_2(q^2)$ the 
chiral-odd part. 

Furthermore, the two-point function of a spin $\frac{3}{2}$ Rarita-Schwinger type 
operator $\eta_{\mu}(x)$ is defined as 
\begin{equation}
\begin{split}
\Pi(q) &= i \int d^{4}x e^{iqx} \langle0|T[\eta_{\mu}(x) \overline{\eta_{\nu}}(0)]|0\rangle \\
&\equiv -g_{\mu\nu}[\Pi_1(q^2) \slash{q} + \Pi_2(q^2)] + \dots, \label{eq:cor2}
\end{split}
\end{equation}
where the dots stand for other Lorentz structures than $g_{\mu\nu}$. In the present study, 
we will only need the terms containing $g_{\mu\nu}$. We use the same notation as for 
the spin $\frac{1}{2}$ case and denote $\Pi_1(q^2)$ as the chiral-even, and $\Pi_2(q^2)$ as the 
chiral-odd part.

In the QCD sum rule approach, we use 
the analytic properties of these two-point functions to extract information of the physical states 
that couple to the operators $\eta$ or $\eta_{\mu}$. 
Concretely, the analyticity of Eqs.(\ref{eq:cor1}) and (\ref{eq:cor2}) allows one to write down the 
dispersion relation 
\begin{equation}
\Pi_i(q^2) = \frac{1}{\pi}\displaystyle \int^{\infty}_{0} ds \frac{\mathrm{Im} \Pi_i(s)}{s-q^2} 
+(\text{subtraction terms}), \label{eq:disp} 
\end{equation} 
for $i=1,2$. This equation has the same form for both the 
spin $\frac{1}{2}$ and spin $\frac{3}{2}$ case. 
To handle a possible divergence in the integral of the right-hand side, usually the subtracted dispersion 
relation is used. This means that subtraction terms have to be added 
to this equation, which contribute mainly to the high-energy part of the spectral function. 
The significance of these terms for the low-energy region is hence small. 
Moreover, the subtraction terms will disappear 
when the Borel transformation is applied as they are polynomials of $q^2$. We will thus omit them in
the following. 

The imaginary part of the two-point function $\mathrm{Im}\Pi_i(s)$, which corresponds to 
the spectral function of $\eta$ or $\eta_{\mu}$, satisfies the following spectral conditions: 
\begin{equation}
\mathrm{Im}\Pi_1(s) \ge 0, \hspace{0.5cm} \sqrt{s}\mathrm{Im}\Pi_1(s) - \mathrm{Im}\Pi_2(s) \ge 0.
\label{eq:positivity}
\end{equation}
Thus, the positivity condition only holds for the chiral-even part, while the spectral function 
obtained from the chiral-odd part is allowed to have negative values. 

In this study, we employ the usual ``pole + continuum" parametrization for the spectral function, which appears in the imaginary 
part of the correlator
\begin{equation}
\mathrm{Im}\Pi_i(s) = \pi(\lambda_i)^2\delta(s - m_{\Theta^+}^2) + \theta(s - s_{th})\mathrm{Im}\Pi_i^{OPE}(s). \label{eq:pole}
\end{equation} 
The $\delta$-function for the ground state pole is justified by the experimental results, which show that the width of 
$\Theta^+(1540)$ is very narrow. The potential contribution of the $KN$ scattering states, which are not 
included in the expression above will be discussed in Sec. \ref{KNscat}. 

Furthermore, to obtain consistent results, it is important to choose an appropriate operator in the two-point function, 
whose spectral function resembles that of Eq.(\ref{eq:pole}) as much as possible. In other words, the chosen operator 
should couple strongly to the ground state pole (if it exists), leading to a large value of the residue $(\lambda_i)$ and 
at the same time should only have a small overlap with the continuum states, which are included in the parametrization 
of Eq.(\ref{eq:pole}) only above the threshold parameter $s_{th}$. In the present study, we will try to construct such an operator by 
considering linear combinations of two independent operators and then fixing the mixing angles so that the results are 
consistent with the ansatz of Eq.(\ref{eq:pole}).    

While the low-energy part of the spectral function below the threshold parameter $s_{th}$ is phenomenologically parametrized as in 
Eq.(\ref{eq:pole}), the left-hand side of Eq.(\ref{eq:disp}) and the second term of Eq.(\ref{eq:pole}) are calculated analytically using the 
OPE. The results of this calculation can be generally expressed as follows:
\begin{equation}
\begin{split}
\Pi^{OPE}_1(q^2) &= \displaystyle \sum^5_{j=0} C_{2j}(q^2)^{5-j}\log(-q^2) + \displaystyle \sum^{\infty}_{j=1} \frac{C_{10+2j}}{(q^2)^j}, \\
\Pi^{OPE}_2(q^2) &= \displaystyle \sum^5_{j=0} C_{2j+1}(q^2)^{5-j}\log(-q^2) + \displaystyle \sum^{\infty}_{j=1} \frac{C_{11+2j}}{(q^2)^j}. \label{eq:ope}
\end{split}
\end{equation}
Here, $C_i$ contain various quark and gluon condensates and numerical factors.

The next step in the calculation is to apply the Borel transformation, which is defined as 
\begin{equation}
L_M[\Pi_i(q^2)] \equiv \lim_{\genfrac{}{}{0pt}{}{-q^2,n \to \infty,}{
-q^2/n=M^2}}
\frac{(-q^2)^{n+1}}{n!} \Bigg(\frac{d}{dq^2}\Bigg)^n 
\Pi_i(q^2), \label{eq:Boreltrans1}
\end{equation}
where $M$ is the so-called Borel mass. There are several reasons for using this transformation: firstly, the high-energy 
continuum part of the spectral function and the higher-order terms in 
the OPE are suppressed by the factors $e^{-s/M^2}$ and $1/n!$, respectively. This considerably improves the accuracy of 
the sum rules. Secondly, as already mentioned above, the Borel transformation removes the subtraction terms in Eq.(\ref{eq:disp}) 
and therefore eliminates possible ambiguities originating from these terms.

Substituting Eq.(\ref{eq:pole}) into the dispersion relation of Eq.(\ref{eq:disp}), and applying the Borel transformation, the following 
expressions can be obtained:
\begin{equation}
\begin{split}
&(\lambda_1)^2 e^{-m_{\Theta^+}^2/M^2} \\
&= -\displaystyle \int_0^{s_{th}} ds e^{-s/M^2} \displaystyle \sum^5_{j=0} C_{2j} s^{5-j} 
+ \displaystyle \sum^{\infty}_{j=1} \frac{(-1)^j C_{10+2j}}{\Gamma(j) (M^2)^{j-1}} \\
&\equiv f_1(M,s_{th}), 
\label{eq:ope1}
\end{split} 
\end{equation}
\begin{equation}
\begin{split}
&(\lambda_2)^2 e^{-m_{\Theta^+}^2/M^2} \\
&= -\displaystyle \int_0^{s_{th}} ds e^{-s/M^2} \displaystyle \sum^5_{j=0} C_{2j+1} s^{5-j}  
+ \displaystyle \sum^{\infty}_{j=1} \frac{(-1)^j C_{11+2j}}{\Gamma(j) (M^2)^{j-1}} \\
&\equiv f_2(M,s_{th}). \\ 
\label{eq:ope2}
\end{split}
\end{equation}
From these equations, the expressions for $m_{\Theta^+}$ and $(\lambda_i)^2$ can be extracted straightforwardly:
\begin{equation}
m_{\Theta^+}^2(M, s_{th}) = \frac{1}{f_i(M,s_{th})} \frac{\partial f_i(M,s_{th})}{\partial(-1/M^2)}, \label{eq:mass}
\end{equation}
\begin{equation}
(\lambda_i)^2 = f_i(M.s_{th})e^{m^2_{\Theta^+}(M,s_{th})/M^2}.
\label{eq:residue}
\end{equation}
Notice that $m_{\Theta^+}$ can be calculated independently either from the chiral-even term 
$f_1(M,s_{th})$ or from the chiral-odd term $f_2(M,s_{th})$. 
In this study we will mainly use $f_1(M,s_{th})$ to calculate $m_{\Theta^+}$ and refer to $f_2(M,s_{th})$ 
only for determining the parity of the investigated state. 

Here, the Borel mass $M$ and the threshold parameter $s_{th}$ 
are variable parameters, which allow us to obtain distinct sum rules in 
Eqs.(\ref{eq:mass}) and (\ref{eq:residue}) for each chosen value of $M$ and $s_{th}$. Comparing 
these different sum rules, it is possible to extract information on the shape of the investigated 
spectral function and on the physical states that contribute to the sum rules most strongly. 
We will discuss this issue in detail in the later sections. 

\subsection{Parity projection}
The parity of the presumed ground state pole 
can not be determined 
from the sum rule of the chiral-even (or chiral-odd) part alone, 
as $\eta(x)$ or $\eta_{\mu}(x)$ couple to both states with positive and negative 
parity regardless of their own intrinsic parity. 
To this end, we use the parity-projected sum rules \cite{Jido} to obtain information 
on the parity of the investigated state. In this method, instead of Eqs.(\ref{eq:cor1}) 
or (\ref{eq:cor2}), the ``old fashioned" Green function is considered 
in the rest frame ($\vec{q}=0$):
\begin{equation}
\begin{split}
\Pi^{of}(q_0) &= i \int d^{4}x e^{iqx} \langle0|\theta(x_0)\eta(x)\overline{\eta}(0)|0\rangle\Big|_{\vec{q}=0} \\
&\equiv \Pi_1^{of}(q_0)\gamma^0 + \Pi_2^{of}(q_0) 
\end{split}
\label{eq:retard}
\end{equation}
Here, only the spin $\frac{1}{2}$ case is shown. 
This expression leads then to two independent sum rules for states coupling to $\eta(x)$ with 
positive and negative parity respectively. These are given as 
\begin{equation}
\begin{split}
&|\lambda_{\pm}|^2e^{-(m_{\Theta^+}^{\pm})^2/M^2} \\
&= \frac{1}{\pi} \int_0^{q_0^{th}} dq_0 \bigl[\mathrm{Im}\Pi^{of}_1(q_0) \pm \mathrm{Im}\Pi^{of}_2(q_0)\bigr]e^{-q_0^2/M^2}, \label{eq:ope3}
\end{split}
\end{equation}
where the intrinsic parity of the operator has been assumed to be positive. In the opposite case the signs of the right-hand 
side have to be switched. 
$q_0^{th}$ is the threshold parameter corresponding to $s_{th}^{1/2}$ in Eq.(\ref{eq:pole}). 

We now have three sum rules, Eqs.(\ref{eq:ope1}),(\ref{eq:ope2}) and (\ref{eq:ope3}), which must in principle 
give the same results for $m_{\Theta^+}$. 
However, the OPE 
of the chiral-odd part $\Pi_2$ has turned out to contain ambiguous terms in the first power of the strange quark 
mass $m_s$, related 
to an infrared divergence originating in the perturbative treatment of $m_s$. To circumvent this problem, 
we will use only the sum rule of the chiral-even part of Eq.(\ref{eq:ope1}) (where the divergencies do not occur) to calculate the 
mass of the ground state. Meanwhile, Eq.(\ref{eq:ope3}) will be applied in the chiral limit 
($\frac{\langle\overline{s}s\rangle}{\langle\overline{q}q\rangle}$ = 1, $m_s = 0$) in order to determine the parity 
of the ground state.

\subsection{Borel window \label{BW}}
It is well known since the QCD sum rule method has been formulated, that the condition of an existing Borel window 
provides an essential check of the accuracy of the method. We define the Borel window as the region of the Borel mass where 
the following two conditions are satisfied.
\begin{equation}
\frac{L_M\bigl[\Pi^{OPE}_{\text{highest order terms}}(q^2)\bigr]}{L_M\bigl[\Pi^{OPE}_{\text{all terms}}(q^2)\bigr]} \le 0.1. \label{eq:conv}
\end{equation}
\begin{equation}
\frac{\displaystyle \int^{s_{th}}_{0}ds e^{-\frac{s}{M^2}}\mathrm{Im} \Pi^{OPE}(s)}
{\displaystyle \int_{0}^{\infty}ds e^{-\frac{s}{M^2}}\mathrm{Im} \Pi^{OPE}(s)}\ge 0.5. \label{eq:polecontr}
\end{equation}
Eq.(\ref{eq:conv}) is a necessary condition for the OPE to converge. It gives a lower limit 
for the Borel mass because the higher-order terms are suppressed for larger $M$ as can be seen in Eqs.(\ref{eq:ope1}) 
and (\ref{eq:ope2}). 
On the other hand, Eq.(\ref{eq:polecontr}) ensures that the low-energy part of the spectral function 
dominates the sum rules, and that contributions from high-energy states do not deteriorate the result.
It is therefore necessary that the high-energy part of the spectral function is sufficiently suppressed. 
As the suppression is stronger for smaller $M$, this condition gives an upper bound for the Borel mass.

One may wonder what the rationale for the numbers on the right-hand side of Eqs.(\ref{eq:conv}) and (\ref{eq:polecontr}) is. 
These numbers are in fact chosen quite reasonably, which can be understood from the following considerations. 
The main uncertainties in a QCD sum rule calculation in most cases 
originate from ambiguities of the vacuum condensates which often have error bars considerably 
larger than $10\,\%$. This justifies the usage of Eq.(\ref{eq:conv}) as choosing any much smaller number than $0.1$ on the 
right-hand side of this condition would be meaningless. 
Moreover, in order for the low-energy states below $s_{th}$ 
to contribute most strongly to the sum rules and that thus an inappropriate parametrization of the high-energy states do not 
introduce too large errors, 
the right-hand side of (\ref{eq:polecontr}) is also a natural choice, as is known from experience 
with sum rules of other baryons and mesons \cite{Reinders}. 
Therefore, one can have some confidence that the errors coming from the 
neglected higher-order terms of the OPE and the possible inaccurate description of the spectral function above $s_{th}$ are 
under control if both conditions of Eqs.(\ref{eq:conv}) and (\ref{eq:polecontr}) are satisfied.

Let us now discuss the difficulties of establishing a Borel window in pentaquark studies. 
As has been discussed above and also in \cite{Matheus2,Kojo1,Gubler}, the existence of a valid Borel window is 
essential for obtaining reliable results within the QCD sum rule technique. Nevertheless almost all earlier studies investigating pentaquark 
states with QCD sum rules \cite{Zhu,Matheus1,Sugiyama,Eidemuller,Ioffe,Kondo,Lee,Nishikawa2} did not consider this problem and 
therefore these results should not be seen as to be conclusive.

The reason why all these studies have ignored this issue, is that, in fact, it is very difficult (if not impossible) to establish a valid 
Borel window in the conventional QCD sum rules described so far in this paper. There are basically two difficulties.  
Firstly, the convergence of the OPE of the correlator 
of a five-quark operator is slower compared to the case of nonexotic baryons containing only three quarks. This can be understood 
from a simple argument: graphs containing quark loops are multiplied by a factor of $(\frac{1}{2\pi})^{2n}$ ($n$: number of quark loops), 
which comes from the integration of the momenta in each loop, and 
the graphs of higher-order terms, where some of the quark loops are cut, are thus enhanced \cite{Ioffe2}. Therefore one can expect 
that the OPE starts to converge only after all the quark loops are being cut. For the pentaquark calculation, this happens only for terms 
of dimension 12 (or higher) and one therefore needs to calculate the OPE at least up to dimension 12 to make sure that all the terms 
with a possible large contribution are included. 
This problem can in principle be solved if one calculates the OPE up to high enough orders (which is a tedious, but straightforward task, 
if one uses the vacuum saturation approximation).

The second difficulty is more severe: because of the high dimensionality of the interpolating field of the pentaquark (15/2 compared to 
9/2 for an ordinary three-quark baryon), the high-energy part of the spectral function well above the presumed $\Theta^+$ resonance 
is enhanced and therefore in many cases dominates the behavior of the whole spectral function. This makes it very difficult to obtain 
a large enough pole contribution in Eq.(\ref{eq:polecontr}) for establishing a valid Borel window. 
It has thus been a very hard task to make a reliable prediction on the resonance $\Theta^+(1540)$.

As discussed in the next section, 
this problem can be solved by a modification of the standard QCD sum rules technique, which consists of using, instead of 
a single correlator, the difference of two independent correlators to construct the sum rules. 
This will be our strategy in this paper.

\section{Details of the method}
\subsection{Interpolating fields}
To carry out QCD sum rule calculations, one first has to construct appropriate operators, which 
carry the desired quantum numbers. These operators should be chosen to 
couple strongly to the state that one wants to investigate, although this is not always 
a trivial task. The interpolating fields that we use in the present study are 
described in this section. Our general strategy is to assemble two independent operators for 
each quantum number and set up general interpolating fields by considering linear combinations of them. 

All the operators used in this study are built from two $ud$ diquarks and an 
$\overline{s}$ antiquark, so that their $KN$ component on the operator level is as small as possible. 
We therefore hope that these operators only have a small overlap with the $KN$ 
scattering states while they should couple strongly to the possible pentaquark resonance. 
For orientation, the properties of the employed $ud$ diquarks are given in 
Table \ref{diquark}. 
\begin{table}
\begin{center}
\caption{The quantum numbers of the $ud$ diquarks used in this study. 
$a, b,\dots$ are color indices and $C=i\gamma^2\gamma^0$ stands for the charge conjugation matrix.} 
\label{diquark}
\begin{tabular}{lccc} \hline
Diquarks \hspace*{0.5cm}& \hspace*{0.5cm} $I$\hspace*{0.5cm}& \hspace*{0.5cm}$J$\hspace*{0.5cm}& 
\hspace*{0.5cm}$\pi$\hspace*{0.5cm} \\ \hline
$\epsilon_{abc}(u^T_a C\gamma_5d_b)$ & $0$ & $0$ & $+$  \\
$\epsilon_{abc}(u^T_a Cd_b)$ & $0$ & $0$ & $-$  \\
$\epsilon_{abc}(u^T_a C\gamma_{\mu} \gamma_5d_b)$ & $0$ & $1$ & $-$  \\
$\epsilon_{abc}(u^T_a C\gamma_{\mu} d_b)$ & $1$ & $1$ & $+$  \\ \hline
\end{tabular}
\end{center}
\end{table}

\subsubsection{The $IJ^P = 0\frac{1}{2}^{\pm}$ and $1\frac{1}{2}^{\pm}$ states}
For the isosinglet case with spin $\frac{1}{2}$, we use the following two operators. 
The same ones were used in \cite{Kojo1}, where the $IJ^P = 0\frac{1}{2}^{\pm}$ states were investigated with a 
similar strategy as in this paper.   
\begin{equation}
\begin{split}
\eta^{I=0}_{1}(x) =
&\epsilon_{cfg}[\epsilon_{abc} u^T_a (x)C\gamma_5d_b (x)] \\
&\times [ \epsilon_{def}u^T_d (x)C\gamma_{\mu} \gamma_5d_e (x)]
\gamma^{\mu}\gamma_5 C\overline{s}^T_g (x),
\label{eq:field1}
\end{split}
\end{equation}
\begin{equation}
\begin{split}
\eta^{I=0}_{2}(x) = 
&\epsilon_{cfg}[\epsilon_{abc}u^T_a (x)Cd_b (x)] \\
&\times [\epsilon_{def}u^T_d (x)C\gamma_{\mu} \gamma_5d_e (x)]
\gamma^{\mu} C\overline{s}^T_g (x).
\label{eq:field2}
\end{split}
\end{equation}
Here, $a, b,\dots$ are color indices, $C$ is the charge conjugation matrix and T stands for the transposition operation. These fields both 
have positive intrinsic parity. They are constructed from a scalar diquark, a vector diquark and an antistrange quark operator in 
the case of $\eta_1$ and from a pseudoscalar diquark, a vector diquark, and an antistrange quark operator in the case of $\eta_2$. 
To project out the spin $\frac{1}{2}$ component, both operators have been multiplied by $\gamma^{\mu}$ and $\eta_1$ is furthermore multiplied by 
$\gamma_5$ to get the correct intrinsic parity. It must be remembered that even though the intrinsic parity of these operators is positive, it can 
couple to both states with positive and negative parity. 

By introducing a mixing angle $\theta^{0}_{1/2}$, a general operator can be constructed from $\eta^{I=0}_1(x)$ and $\eta^{I=0}_2(x)$:
\begin{equation}
\eta^{I=0}(x)=\cos\theta^{0}_{1/2}\eta^{I=0}_{1}(x)+\sin\theta^{0}_{1/2}\eta^{I=0}_{2}(x). \label{eq:mix1}
\end{equation}
This is the operator that will be used in the actual calculation. Here, we are in principle allowed to choose any value for the 
mixing angle. The strategy for determining this free parameter will be given in the following subsection.

For the isotriplet, we employ the following two operators, which have a form similar to the isosinglet case:
\begin{equation}
\begin{split}
\eta^{I=1}_{1}(x) =
&\epsilon_{cfg}[\epsilon_{abc} u^T_a (x)C\gamma_5d_b (x)] \\
&\times [ \epsilon_{def}u^T_d (x)C\gamma_{\mu} d_e (x)]
\gamma^{\mu}C\overline{s}^T_g (x),
\label{eq:field3}
\end{split}
\end{equation}
\begin{equation}
\begin{split}
\eta^{I=1}_{2}(x) = 
&\epsilon_{cfg}[\epsilon_{abc}u^T_a (x)Cd_b (x)] \\
&\times [\epsilon_{def}u^T_d (x)C\gamma_{\mu} d_e (x)]
\gamma^{\mu} \gamma_5 C\overline{s}^T_g (x).
\label{eq:field4}
\end{split}
\end{equation} 
The notation is the same as before. The difference to the fields with $I=0$ is that instead of a vector diquark, we here have an 
axial-vector diquark, which carries isospin $I=1$. Moreover, note that both the operators are multiplied by $(-\gamma_5)$ to obtain 
positive intrinsic parity for $\eta^{I=1}_1(x)$ and $\eta^{I=1}_2(x)$. Analogously to the isosinglet case, we construct a general 
operator by introducing a mixing angle $\theta^{1}_{1/2}$:
\begin{equation}
\eta^{I=1}(x)=\cos\theta^{1}_{1/2}\eta^{I=1}_{1}(x)+\sin\theta^{1}_{1/2}\eta^{I=1}_{2}(x). \label{eq:mix2}
\end{equation}

\subsubsection{The $IJ^P = 0\frac{3}{2}^{\pm}$ and $1\frac{3}{2}^{\pm}$ states}
The construction of the operators with spin $\frac{3}{2}$ can be done in a similar fashion as for spin $\frac{1}{2}$. There are, 
however, some additional steps arising from the properties of the spin $\frac{3}{2}$ Rarita-Schwinger type fields. These are, for instance, 
discussed in \cite{Hwang,Gubler} and we do not repeat the details here. We only state the result of how the 
spin $\frac{3}{2}$ components can be extracted. In the case of Rarita-Schwinger fields, 
the two-point function of Eq.(\ref{eq:cor2}) 
generally contains various different tensor structures, with contributions from states with 
spin $\frac{1}{2}$ and $\frac{3}{2}$. It can be shown that the terms proportional to $g_{\mu\nu}$ receive only contributions 
from the spin $\frac{3}{2}$ states. Therefore, if one considers only the two terms
\begin{equation}
\Pi_{\mu\nu}(q) = -g_{\mu\nu}[\Pi_1(q^2)\slash{q} + \Pi_2(q^2)] + \dots \label{eq:evenodd}
\end{equation}
the spin $\frac{1}{2}$ contributions will automatically be eliminated. Note that there is a minus sign on the right-hand side of 
Eq.(\ref{eq:evenodd}), which is a consequence of the properties of the Rarita-Schwinger field. 

To study the isosinglet states we employ the following interpolating fields
\begin{equation}
\begin{split}
\eta^{I=0}_{1,\mu}(x) =
&\epsilon_{cfg}[\epsilon_{abc} u^T_a (x)C\gamma_5d_b (x)] \\
&\times [ \epsilon_{def}u^T_d (x)C\gamma_{\mu} \gamma_5d_e (x)]
C\overline{s}^T_g (x),
\label{eq:field5}
\end{split}
\end{equation}
\begin{equation}
\begin{split}
\eta^{I=0}_{2,\mu}(x) = 
&\epsilon_{cfg}[\epsilon_{abc}u^T_a (x)Cd_b (x)] \\
&\times [\epsilon_{def}u^T_d (x)C\gamma_{\mu} \gamma_5d_e (x)]
\gamma_5C\overline{s}^T_g (x).
\label{eq:field6}
\end{split}
\end{equation}
These operators have the same structure as the ones with spin $\frac{1}{2}$ [Eqs.(\ref{eq:field1}) and (\ref{eq:field2})]. The only 
difference is that $\gamma^{\mu}\gamma_5$in front of $C \overline{s}^T$ has been omitted here, which allows the operators 
to couple to spin $\frac{3}{2}$ states and lets the intrinsic parity become positive. As above, a general operator is then 
constructed by a linear combination of  $\eta^{I=0}_{1,\mu}$ and $\eta^{I=0}_{2,\mu}$:
\begin{equation}
\eta^{I=0}_{\mu}(x)=\cos\theta^{0}_{3/2}\eta^{I=0}_{1,\mu}(x)+\sin\theta^{0}_{3/2}\eta^{I=0}_{2,\mu}(x). \label{eq:mix3}
\end{equation}
This is the same kind of operator that has been 
used in our previous work \cite{Gubler}. In this paper we will merely restate the results that have been obtained there in order to 
compare them with the results from the other quantum numbers.

Finally, for the isotriplet case, we will use the operators given below:
\begin{equation}
\begin{split}
\eta^{I=1}_{1,\mu}(x) =
&\epsilon_{cfg}[\epsilon_{abc} u^T_a (x)C\gamma_5d_b (x)] \\
&\times [ \epsilon_{def}u^T_d (x)C\gamma_{\mu} d_e (x)]
\gamma_5 C\overline{s}^T_g (x),
\label{eq:field7}
\end{split}
\end{equation}
\begin{equation}
\begin{split}
\eta^{I=1}_{2,\mu}(x) = 
&\epsilon_{cfg}[\epsilon_{abc}u^T_a (x)Cd_b (x)] \\
&\times [\epsilon_{def}u^T_d (x)C\gamma_{\mu} d_e (x)]
C\overline{s}^T_g (x).
\label{eq:field8}
\end{split}
\end{equation}
The structure of these operators is almost the same as the ones with quantum numbers $IJ^P=1\frac{1}{2}^{\pm}$. 
Here again, compared with Eqs.(\ref{eq:field3}) and (\ref{eq:field4}) the matrices $\gamma^{\mu} \gamma_5$ have been 
omitted in order to construct Rarita-Schwinger fields which couple to spin $\frac{3}{2}$ states and to adjust the 
intrinsic parity to be positive. As in all the cases above, a general operator $\eta^{I=1}_{\mu}$ is constructed from 
$\eta^{I=1}_{1,\mu}$ and $\eta^{I=1}_{2,\mu}$, which will then be used to formulate the sum rules
\begin{equation}
\eta^{I=1}_{\mu}(x)=\cos\theta^{1}_{3/2}\eta^{I=1}_{1,\mu}(x)+\sin\theta^{1}_{3/2}\eta^{I=1}_{2,\mu}(x). \label{eq:mix4}
\end{equation}

\subsection{Determination of the Borel mass, threshold parameter and mixing angle \label{detpar}}
The Borel mass $M$ appears in the formulation of QCD sum rules when the Borel transformation is applied in Eq.(\ref{eq:Boreltrans1}), 
the threshold parameter $s_{th}$ in the ``pole+continuum" ansatz of Eq.(\ref{eq:pole}), and the mixing angles $\theta^I_{J}$ in the general 
expressions for the interpolating fields in Eqs.(\ref{eq:mix1}),(\ref{eq:mix2}),(\ref{eq:mix3}) and (\ref{eq:mix4}). 
In this subsection, our strategy of determining these parameters will be explained.

Let us first discuss the question of how the Borel mass $M$ has to be determined. As mentioned in the last section, it first 
has to be checked whether one can establish a valid Borel window from the sum rules. If not, the sum rules will not work and 
it will not be possible to obtain any reliable results from them. If one is able to find a valid Borel window, $M$ has to be 
chosen within its boundaries. As will be discussed below, when the sum rules ``work well", the dependence of the results on
$M$ should be small and therefore it will not strongly depend on the exact position on $M$ inside of the Borel window.

Next, our strategy of determining the threshold parameter $s_{th}$ will be 
explained. Assuming that the low-energy part of the spectral function is dominated by a narrow resonance pole, 
the values of the resonance mass [given in Eq.(\ref{eq:mass})] and the residue [Eq.(\ref{eq:residue})] should 
not strongly depend on $M$ and $s_{th}$. This is easily understood when one considers the (ideal) case, when the 
spectral function is given by a single $\delta$-function below $s_{th}$. Rewriting the right-hand side 
of Eq.(\ref{eq:mass}), we obtain 
\begin{equation}
\begin{split}
& \frac{\frac{\partial}{\partial (-1/M^2)}\displaystyle \int_{0}^{s_{th}}
ds e^{-s/M^2} \mathrm{Im\Pi(s)}  }{\displaystyle \int_{0}^{s_{th}}
ds e^{-s/M^2} \mathrm{Im\Pi(s)} } = \\
& \frac{\displaystyle \int_{0}^{s_{th}}
ds e^{-s/M^2} s\mathrm{Im\Pi(s)}}
{\displaystyle \int_{0}^{s_{th}}
ds e^{-s/M^2} \mathrm{Im\Pi(s)} }.
\label{eq:dependence}
\end{split}
\end{equation}
If $\mathrm{Im\Pi(s)}$ is a simple $\delta$-function specified as 
$\mathrm{Im}\Pi(s) = \pi(\lambda)^2\delta(s - m^2) + \theta(s - s_{th})\mathrm{Im}\Pi^{'}(s)$, 
then Eq.(\ref{eq:dependence}) gives $m^2$ and 
does not depend on $M$ and $s_{th}$. On the other hand, 
if $\mathrm{Im\Pi(s)}$ is described by some continuous positive curve, which corresponds to the scattering states, 
Eq.(\ref{eq:dependence}) should be a rising curve, because of the weight factor $e^{-s/M^2}$, which suppresses the 
part of the integral with large $s$ values when $M$ is small. Furthermore, Eq.(\ref{eq:dependence}) should 
have an increasing value when $s_{th}$ is raised, as higher values of $s$ will be included in the integral. 

Following the arguments above, it can be understood that the threshold parameter $s_{th}$ 
has to be chosen so that the dependence of the calculated resonance mass and its residue is 
smallest, because this corresponds to the case of the largest contribution of a narrow ground state pole to the 
spectral function. On the other hand, if no such value for $s_{th}$ can be found, we can assume that 
the spectral function is dominated by the scattering states.

We therefore set up the following two conditions, by which we determine $s_{th}$ 
(called in the following the conditions of pole domination): 
\begin{description}
\item[1)] A sufficiently wide Borel window exists.
\item[2)] $m_{\Theta^+}(M,s_{th})$ should only weakly depend on the Borel mass $M$ and on the threshold parameter $s_{th}$.
\end{description}
Condition 1) is essential to obtain reliable results with the QCD sum rule method, while 2) follows from the discussion above. 
The problem that arises here, is how we should quantitatively define the ``weak dependence" of condition 2). In other words, 
how ``weak" should the dependence on $M$ and $s_{th}$ be that one can be unambiguously sure not just to observe 
scattering states? 
This important problem will be discussed in the part of the result section, which deals with $KN$ scattering states.

Finally, the mixing angle $\theta^I_J$ has to be fixed. To do this, we repeat the analysis outlined above for 
various values of $\theta^I_J$ and at the end choose the one for which the conditions 
1) and 2) are best 
satisfied. This concludes our discussion about the determination of the different parameters 
that appear in the sum rules.  

\subsection{Establishment of a valid Borel window}
As was pointed out in Sec. \ref{BW}, it has so far been very difficult to establish a valid Borel window in QCD sum rule studies 
of pentaquarks.   
We aim to solve 
this problem by a modification of the standard QCD sum rules technique \cite{Kojo1}. The idea is to use, instead of 
a single correlator, the difference of two independent correlators to construct the sum rules. By this trick, it is hoped that 
we will achieve a large cancellation of the high-energy part of the spectral function, due to the restored chiral symmetry 
in this region. 
We will then be able to obtain a large pole contribution, which (if the OPE
is calculated up to a sufficiently high dimension) will make it possible to establish 
a valid Borel window. 

To illustrate this point more concretely, let us consider the difference of two independent correlators that have been constructed in the 
second part of this section. We take as an example the operators with quantum numbers $IJ^P = 0\frac{1}{2}^{\pm}$. In the same way 
as it was shown in \cite{Gubler} for operators of spin $\frac{3}{2}$, the operators of Eqs.(\ref{eq:field1}) and (\ref{eq:field2}) can be 
expressed with the help of the operators $\xi_1$ and $\xi_2$, which belong to different chiral multiplets:
\begin{equation}
\begin{split}
\xi_{1} 
=&  -2(u^T_R Cd_R)[(u^T_LC\gamma_{\mu}d_R) - (u^T_RC\gamma_{\mu}d_L)]\gamma^{\mu}C\overline{s}^T_R \\
&- 2(u^T_L Cd_L)[(u^T_LC\gamma_{\mu}d_R) - (u^T_RC\gamma_{\mu}d_L)]\gamma^{\mu}C\overline{s}^T_L, \\ 
\xi_{2} 
=&  2(u^T_R Cd_R)[(u^T_LC\gamma_{\mu}d_R) - (u^T_RC\gamma_{\mu}d_L)]\gamma^{\mu}C\overline{s}^T_L \\
&+ 2(u^T_L Cd_L)[(u^T_LC\gamma_{\mu}d_R) - (u^T_RC\gamma_{\mu}d_L)]\gamma^{\mu}C\overline{s}^T_R.
\end{split} \label{eq:chiral}
\end{equation}
Here, the color indices are omitted for simplicity. $\xi_{1}$ belongs 
to the $(\mathbf{3},\overline{\mathbf{15}}) \oplus (\overline{\mathbf{15}},\mathbf{3})$ multiplet 
of the chiral $\mathrm{SU}(3)_L \otimes \mathrm{SU}(3)_R$ group 
with 4(1) right-handed and 1(4) left-handed quarks, 
and $\xi_{2}$ to the $(\mathbf{8},\mathbf{8})$ multiplet with 3(2) right-handed and 2(3) left-handed quarks.

Using these chiral operators, $\eta^{I=0}_{1}$ and $\eta^{I=0}_{2}$ are given as 
\begin{equation}
\begin{split}
\eta^{I=0}_1 &= \frac{1}{2}(\xi_1 + \xi_2), \\
\eta^{I=0}_2 &= \frac{1}{2}(\xi_1 - \xi_2).
\end{split} 
\label{eq:chiralexpression}
\end{equation}
Thus, the correlator of the general operator of Eq.(\ref{eq:mix1}), denoted as 
$\Pi^0_{1/2}(q^2,\theta^0_{1/2}) \equiv \langle \eta^{I=0} \overline{\eta^{I=0}}\rangle$, can be expressed as follows:
\begin{equation}
\begin{split}
\Pi^0_{1/2}(q^2,\theta^0_{1/2}) =& \frac{1}{4} [\langle \xi_1 \overline{\xi_1} \rangle + \langle \xi_2 \overline{\xi_2} \rangle] \\
& +\frac{1}{4}\cos(2\theta^0_{1/2}) [\langle \xi_1 \overline{\xi_2} \rangle + \langle \xi_2 \overline{\xi_1} \rangle] \\
& +\frac{1}{4}\sin(2\theta^0_{1/2}) [\langle \xi_1 \overline{\xi_1} \rangle - \langle \xi_2 \overline{\xi_2} \rangle].
\end{split} 
\label{eq:correlatorxi}
\end{equation}
The first term of this expression does not depend on the mixing angle $\theta^0_{1/2}$, but is expected to couple strongly 
to the high-energy continuum states because this term can have perturbative parts. 
On the other hand, compared with 
the first term, the coupling to the high-energy states of the other two terms 
is expected to be smaller, which can be understood from the following arguments. 
The perturbative part of the second of Eq.(\ref{eq:correlatorxi}) term vanishes because $\xi_1$ and $\xi_2$ belong to 
different chiral multiplets and therefore at least one nonperturbative quark condensate related to 
chiral symmetry breaking is needed to connect them. 
As the perturbative term largely couples to the high-energy states, their contributions will be suppressed in this term. 
Considering the third term, 
it is possible to cancel the leading perturbative terms with an appropriate normalization of 
$\xi_1$ and $\xi_2$. 
Note that we here have implicitly used the positivity condition of the spectral function. As is seen in Eq.(\ref{eq:positivity}), 
this assumption is not necessarily valid for the chiral-odd part, but we employ in this paper only the sum rule 
of the chiral-even part and the parity-projected sum rules, where the positivity condition holds.

Therefore, by taking the difference of two correlators with different mixing angles $\theta^0_{1/2}$ and $\theta^{0'}_{1/2}$, 
the first term in Eq.(\ref{eq:correlatorxi}) will be eliminated, and a strong suppression of the high-energy continuum part 
can be obtained. It will thus become possible to establish a valid Borel window. Writing this difference down, we get 
\begin{equation}
\begin{split}
& \Pi^0_{1/2}(q^2,\theta^0_{1/2}) - \Pi^0_{1/2}(q^2,\theta^{0'}_{1/2}) = \\
& \frac{1}{2}\sin(\theta^0_{1/2} -\theta^{0'}_{1/2}) 
\Bigl\{ \cos(\theta^0_{1/2} + \theta^{0'}_{1/2}) [\langle \xi_1 \overline{\xi_1} \rangle - \langle \xi_2 \overline{\xi_2} \rangle] \\
& -\sin(\theta^0_{1/2} + \theta^{0'}_{1/2})  [\langle \xi_1 \overline{\xi_2} \rangle + \langle \xi_2 \overline{\xi_1} \rangle] \Bigr\},
\end{split} 
\label{eq:diff}
\end{equation}
which will be used to formulate the sum rules. It is understood that the factor $\sin(\theta^0_{1/2} -\theta^{0'}_{1/2})$ has no 
influence on the mass of the ground state, calculated in Eq.(\ref{eq:mass}). We hence fix it at 
$\theta^0_{1/2} -\theta^{0'}_{1/2} = \frac{\pi}{2}$ and will only keep $\theta^0_{1/2} +\theta^{0'}_{1/2} \equiv \phi^0_{1/2}$ as 
a free parameter, which will have to be determined by the conditions stated in the last subsection.

\subsection{Possible contribution of $KN$ scattering states \label{KNscat}}
We have in this paper several times mentioned the possible influence of the $KN$ scattering states 
to the sum rules. 
Generally, if such scattering states have the same quantum numbers as the interpolating fields, they 
may always contribute to the sum rules to a certain extent, so we have to find a way to distinguish them from narrow 
pole states that we are really interested in. We have already mentioned in the discussion of Eq.(\ref{eq:dependence}), 
that in the ideal case when only one narrow pole is present in the spectral function, the results of the sum rules 
should not depend on the Borel mass $M$ and the threshold parameter $s_{th}$. When only scattering states contribute 
to the spectral function, this behavior should change. The nature and extent of this change will be illustrated in this 
section.

Let us first consider how the contribution of the $KN$ scattering states to the spectral function should look like. 
It is known that the $KN$ interaction in the $S=+1$ channel is weak and slightly repulsive for $I=0$, 
while the repulsion is stronger for $I=1$ \cite{Hashimoto,Gibbs}. 
As an illustration, we will here use phase space as a first 
approximation of the $KN$ spectral function, which thus corresponds more closely to the $I=0$ case. 
Nevertheless, the qualitative behavior of the results of this section does not strongly depend on the 
detailed form of the spectral function and can therefore be considered to be quite general. 

In the case of spin $\frac{1}{2}$ states, the contribution of $KN$ phase space to the spectral function can be 
expressed as follows:
\begin{equation}
\begin{split}
& \rho(s=q^2)=\frac{1}{\pi}|\lambda_{KN}|^2 \mathrm{Im}\Biggl\{(-i)
\displaystyle \int\frac{d^4p}{(2\pi)^4} \times \\
& \frac{1}{(q-p)^2-m^2+i\epsilon}\gamma_5\frac{\slash{p}+M}{p^2-M^2+i\epsilon}(-\gamma_5)
\Biggr\}.
\end{split}
\end{equation}
Here, $\lambda_{KN}$ is the coupling of the used interpolating field to the $KN$ scattering states. 
$m$ and $M$ are the masses of the kaon and the nucleon, respectively. The $\gamma_5$`s are necessary 
because of the p-wave coupling of our interpolating field to the nucleon and kaon fields. 

Going to the rest frame ($\vec{q}=0$), 
the integral can be easily calculated and one gets
\begin{equation}
\rho(q_0^2)=\frac{1}{4\pi^2}|\lambda_{KN}|^2\frac{\sqrt{E_N^2-M^2}}{4q_0}(\gamma_0E_N-M), 
\end{equation}
where $E_N$ is the energy of the nucleon, expressed as
\begin{equation}
E_N=\frac{q_0^2+M^2-m^2}{2q_0}.
\end{equation}
Therefore, the spectral function for the chiral-even part has the form 
\begin{equation}
\frac{1}{\pi}\mathrm{Im}\Pi_1^{KN}(q_0^2)=\frac{1}{4\pi^2}|\lambda_{KN}|^2\frac{E_N\sqrt{E_N^2-M^2}}{4q_0^2},
\label{eq:spin1/2}
\end{equation}
which contains contributions from both positive and negative parity states.

For spin $\frac{3}{2}$ states, similar considerations can be applied, although there are some complications 
coming from projecting out the contributions of the spin $J=\frac{3}{2}$ states from the correlator. 
In this case, the spectral function is expressed as
\begin{equation}
\begin{split}
& \rho(s=q^2)=\frac{1}{\pi}\frac{|\lambda'_{KN}|^2}{m^2}\mathrm{Im}\Biggl\{(-i)
\displaystyle \int\frac{d^4p}{(2\pi)^4} \times \\ 
& \frac{(q-p)_{\mu}(q-p)_{\nu}}{(q-p)^2-m^2+i\epsilon}\frac{\slash{p}+M}{p^2-M^2+i\epsilon}
\Biggr\}\Bigg|_{J=\frac{3}{2}}, 
\end{split}
\end{equation}
and the $J=\frac{3}{2}$ projection is most easily done by applying the projection operator 
\begin{equation}
P_{\mu\nu}(q) = g_{\mu\nu} - \frac{2q_{\mu}q_{\nu}}{3q^2} -\frac{1}{3}\gamma_{\mu}\gamma_{\nu} -
\frac{1}{3q^2}(q_{\mu}\gamma_{\nu} - q_{\nu}\gamma_{\mu})\slash{q}.
\end{equation}
Then, only considering the terms proportional to $g_{\mu\nu}$ and again going to the 
rest frame, one obtains 
\begin{equation}
\begin{split}
\rho(q_0^2)= & \frac{1}{12\pi^2m^2}|\lambda'_{KN}|^2\frac{(E_N^2-M^2)^{3/2}}{4q_0}g_{\mu\nu}(\gamma_0E_N+M) \\
&+ \dots, 
\end{split} 
\end{equation}
from which finally the spectral function of the chiral-even part can be extracted:
\begin{equation}
\frac{1}{\pi}\mathrm{Im}\Pi_1^{KN}(q_0^2)=\frac{1}{12\pi^2m^2}|\lambda'_{KN}|^2\frac{E_N(E_N^2-M^2)^{3/2}}{4q_0^2}.
\label{eq:spin3/2}
\end{equation}
This expression again contains contributions from both positive and negative parity.

\begin{figure*}
\includegraphics[width=8cm,clip]{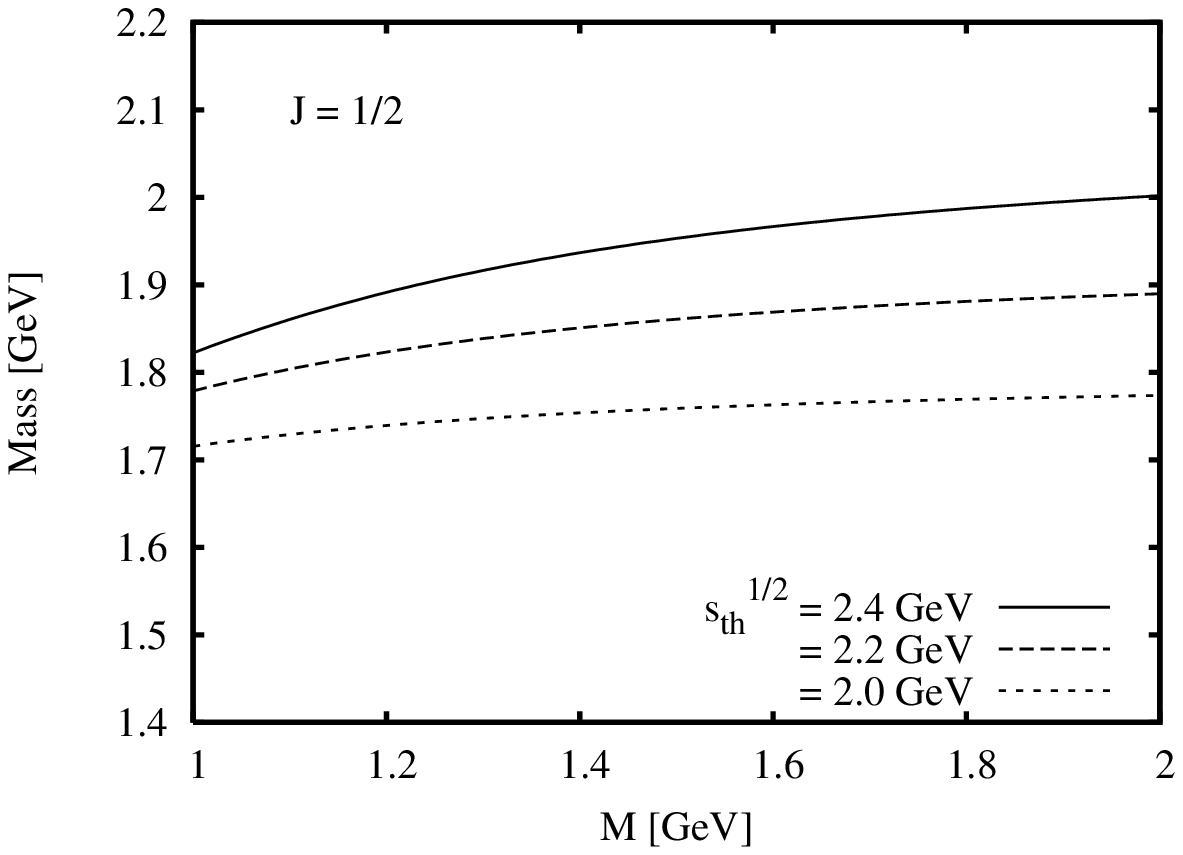}
\includegraphics[width=8cm,clip]{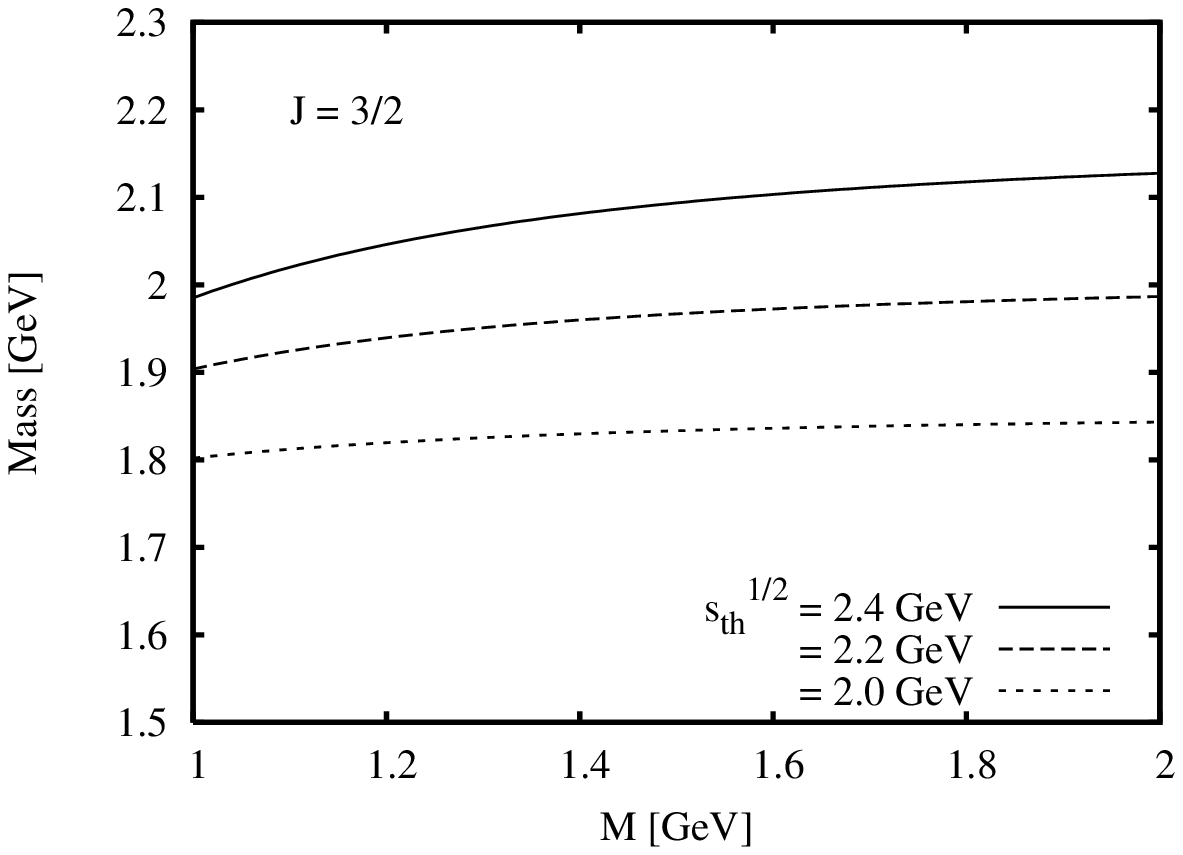}
\caption{The ``mass of the ground state pole" of spin $\frac{1}{2}$ (left) and $\frac{3}{2}$ (right), 
obtained when only the $KN$ scattering states 
contribute to the spectral function. Eq.(\ref{eq:spin1/2}) and 
Eq.(\ref{eq:spin3/2}) have been used as the expression of the spectral function 
of the chiral even-part.}
\label{fig:KNspin0.5}
\end{figure*}

Next, we compute the results that would be obtained by the QCD sum rules 
if only the $KN$ scattering states contribute to the spectral function. This means that we 
calculate the quantity corresponding to Eq.(\ref{eq:mass}) or Eq.(\ref{eq:dependence}), where for 
$\Pi(s)$, we now use the expressions obtained above. The results for spin $\frac{1}{2}$ and spin 
$\frac{3}{2}$ are given in Fig. \ref{fig:KNspin0.5}.

It is clearly seen that while the dependence on the Borel mass $M$ is relatively weak, the results depend strongly on the 
threshold parameter $s_{th}$. This can intuitively be understood from the fact that the spectral function containing only the phase space 
contribution is a fastly growing function with increasing energy. Therefore, the high-energy regions below the threshold parameter 
will dominate the integral of Eq.(\ref{eq:dependence}), which then leads to a behavior as seen in Fig. \ref{fig:KNspin0.5}, 
with a strong dependence on $s_{th}$. Furthermore, this dominance of the high-energy states will make it difficult to obtain a 
large value for the pole contribution and to establish a valid Borel window. 

Note that we have here assumed that $\lambda_{KN}$ and $\lambda'_{KN}$ to be constants with dimension $[\mathrm{GeV}^{5}]$. They could 
in principle also have a dependence on $q_0$ such as $\mathrm{const.} \times q_0^5$, which would result in Eqs.(\ref{eq:spin1/2}) and (\ref{eq:spin3/2}) 
being multiplied by $q_0^{10}$. We have checked this case, finding that the results are altered quantitatively, namely that the mass is shifted 
upwards while the difference between the different threshold curves increases. Nevertheless, our quantitative discussion above does not have to 
be changed, as our observation of a large dependence of the results on $s_{th}$ is even more apparent in this case.    

The results of this section show that the dependence of $m_{\Theta^+}^2(M,s_{th})$ on $s_{th}$ provides us with an indicator of how much the 
$KN$ scattering states contribute to the sum rule: a linear dependence of the same (or larger) extent as in Fig. \ref{fig:KNspin0.5} 
suggests a strong contribution of the scattering states, while a significantly smaller dependence indicates that a narrow pole exists and is the 
dominant structure in the spectral function. Similar arguments have already been discussed earlier in \cite{Kojo2}.

\section{Results}
\subsection{Preliminaries}
We summarize in this subsection general issues common to the sum rules of all the various quantum numbers, 
and explain the parameters, conventions and basic approximations used in the calculation. 

One important feature of the results for all quantum numbers is, that the perturbative term $C_0$ vanishes 
when the difference of the two correlators is taken. This corresponds to the 
suppression of the contribution of the high-energy states as was discussed in the last section. 
The detailed results of the OPE of both the chiral-even and chiral-odd part are given in Appendix \ref{apA}.

We will for all quantum numbers first investigate the sum rule for the chiral-even part and 
after that consider the parity-projected sum rules, where both 
the results of the chiral-even and chiral-odd parts have to be used. 
However, as mentioned before, the results of the OPE calculation 
of the chiral-odd part have turned out to contain some ambiguous terms in the first power of the strange quark 
mass $m_s$, related to an infrared divergence. 
It is important to note here that this kind of divergence is artificially arising because we are expanding our results 
in $m_s$ and are ignoring higher-order terms. It should thus in principle be possible to remove this 
divergence by taking the full dependence on $m_s$ into account without resorting to any expansion, 
although this may be difficult to achieve in practice. 
In any case, to avoid this kind of artificial ambiguity, 
the results of the parity-projected sum rules will be given only in the chiral limit 
($m_s=0$, $\langle\overline{s}s\rangle = \langle\overline{q}q\rangle$). 
Hence, our strategy will be to use the results of the chiral-even part to calculate 
the mass of the investigated state, while we will employ the parity-projected sum 
rules to determine only the parity of the state. 

The values of the mixing angles and threshold parameters are obtained using 
the conditions of pole domination of Sec. \ref{detpar}. We will use the same values 
for both the sum rule of the chiral-even part and the parity-projected sum rules.  

The values of the condensates and other used parameters are given in Table \ref{parameters}.
\begin{table}
\begin{center}
\caption{Values of all the parameters used throughout this paper, 
given at a scale of $1\,\mathrm{GeV}$ \cite{Colangelo,Reinders}. 
The parameter $\kappa$ describes the possible breaking of the vacuum saturation approximation and 
is explained in this subsection.} 
\label{parameters}
\begin{tabular}{lc}
\hline
$\langle\overline{q}q\rangle$ & $-(0.23\pm0.02\,\,\mathrm{GeV})^3$ \\
$\frac{\langle\overline{s}s\rangle}{\langle\overline{q}q\rangle}$ & $0.8\pm0.2$ \\
$\frac{\langle\overline{q}g \sigma\cdot G q\rangle}{\langle\overline{q} q \rangle}$ & $0.8\pm0.1 \,\,\mathrm{GeV}^2$ \\
$\frac{\langle\overline{s}g \sigma\cdot G s\rangle}{\langle\overline{s} s \rangle}$ & $0.8\pm0.1 \,\,\mathrm{GeV}^2$ \\
$\langle\frac{\alpha_s}{\pi}G^2\rangle$ & $0.012\pm0.004\,\,\mathrm{GeV}^4$ \\ 
$m_s$ & $0.12\pm0.06\,\,\mathrm{GeV}$ \\
$\kappa$ & $1\sim2$ \\
\hline
\end{tabular}
\end{center}
\end{table}
These are standard values for QCD sum rule calculations \cite{Colangelo,Reinders}, 
but they of course all have a certain range and the 
results will therefore depend on what exact values have been chosen for the condensates and other parameters. 
In the last part of this result section we will show to what extent the results will be influenced by the uncertainties 
of these parameters.

Finally, $\kappa$, the last parameter of Table \ref{parameters}, 
will now be explained. It parametrizes the possible violation of the vacuum saturation approximation 
and is used as follows:
\begin{equation}
\begin{split}
\langle\overline{q}q \overline{q}q\rangle &= \kappa \langle\overline{q}q\rangle^2, \\
\langle\overline{q}q \overline{q}q\overline{q}q\rangle &= \kappa^2 \langle\overline{q}q\rangle^3, \\
\langle\overline{q}q \overline{q}g\sigma \cdot Gq\rangle &= \kappa \langle\overline{q}q\rangle\langle\overline{q}g\sigma \cdot Gq\rangle, \\
\langle\overline{q}g\sigma \cdot Gq \overline{q}g\sigma \cdot Gq\rangle &= \kappa \langle\overline{q}g\sigma \cdot Gq\rangle^2, \\
&\dots
\end{split}
\end{equation}
All the results shown below are obtained with $\kappa = 1$, which means that the vacuum saturation approximation has been 
assumed. 
This approximation has been shown to be valid in the leading order of the large $N_c$ expansion, even though the $\frac{1}{N_c}$ corrections 
may be quite large. 
We have checked to what extent the results change when this approximation is broken up to values of $\kappa = 2$. These 
changes will be included in the estimation of the error.

The results of the various sum rules are given in the following. To allow a direct 
comparison between the different quantum numbers, all the plots corresponding to the 
same quantity are shown in the same figure.

\subsection{$IJ^{\pi} = 0\frac{1}{2}^{\pm}$}

\begin{figure*}
\begin{center}
\includegraphics[width=8cm,clip]{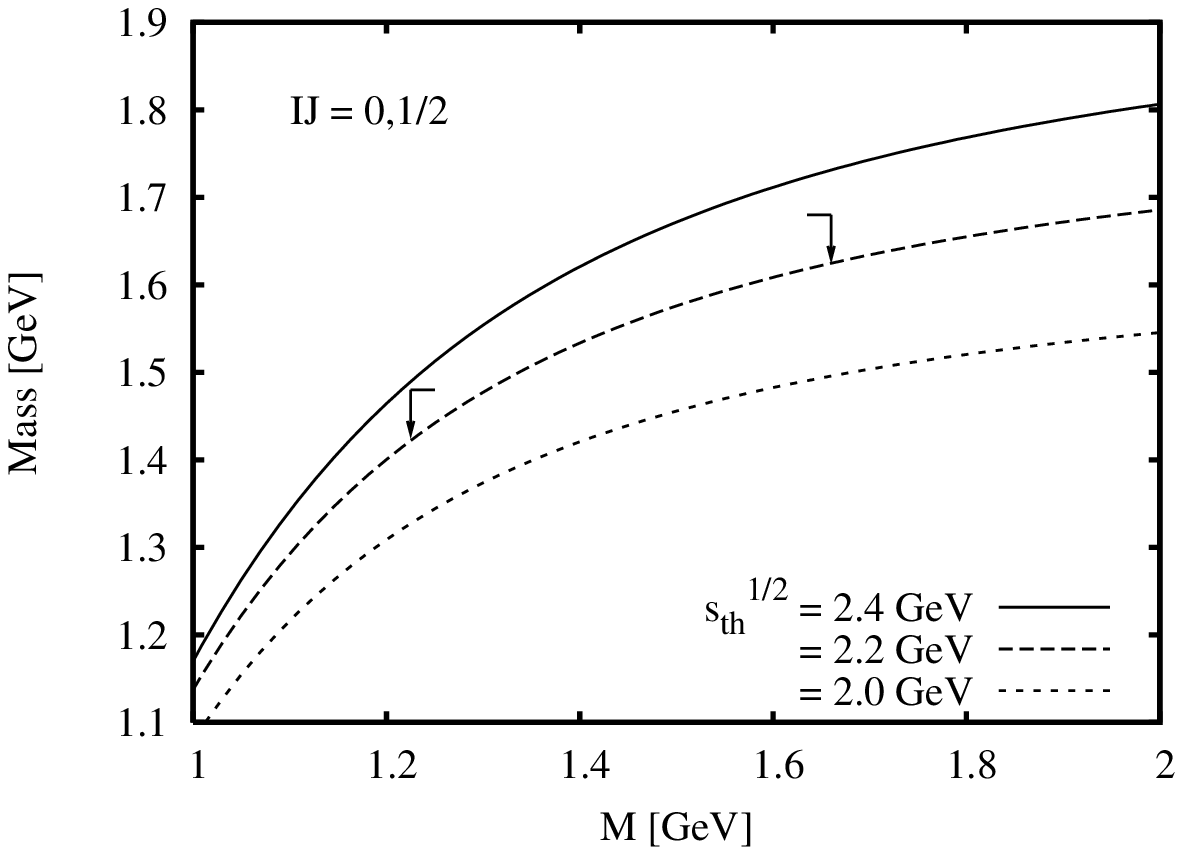} 
\includegraphics[width=8cm,clip]{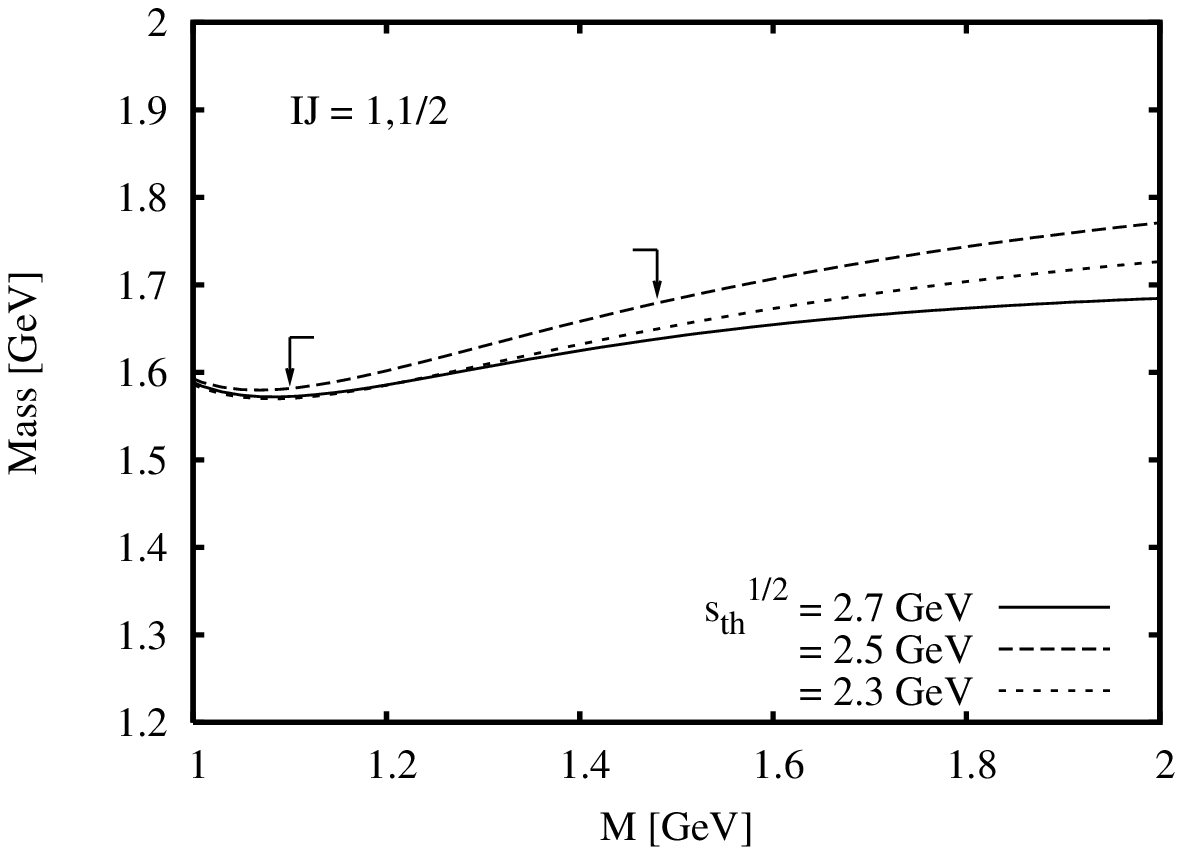} 
\includegraphics[width=8cm,clip]{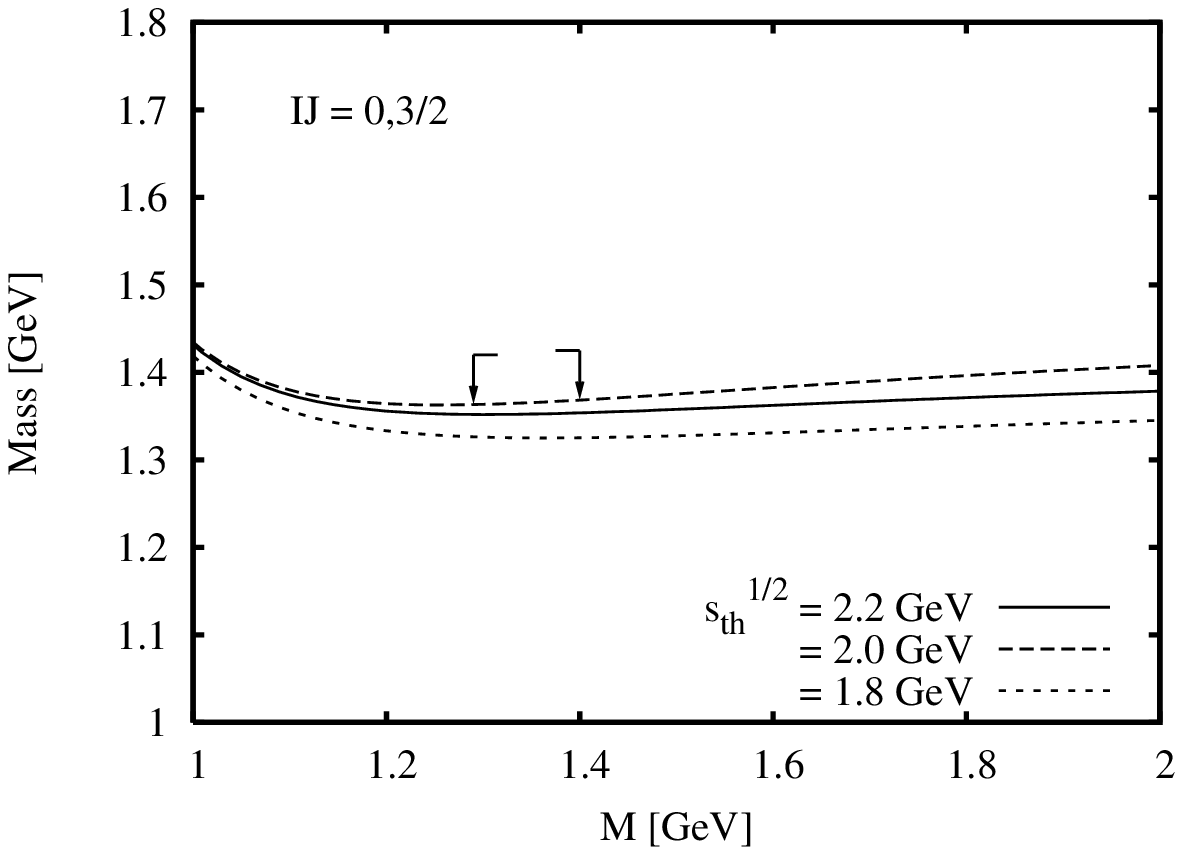} 
\includegraphics[width=8cm,clip]{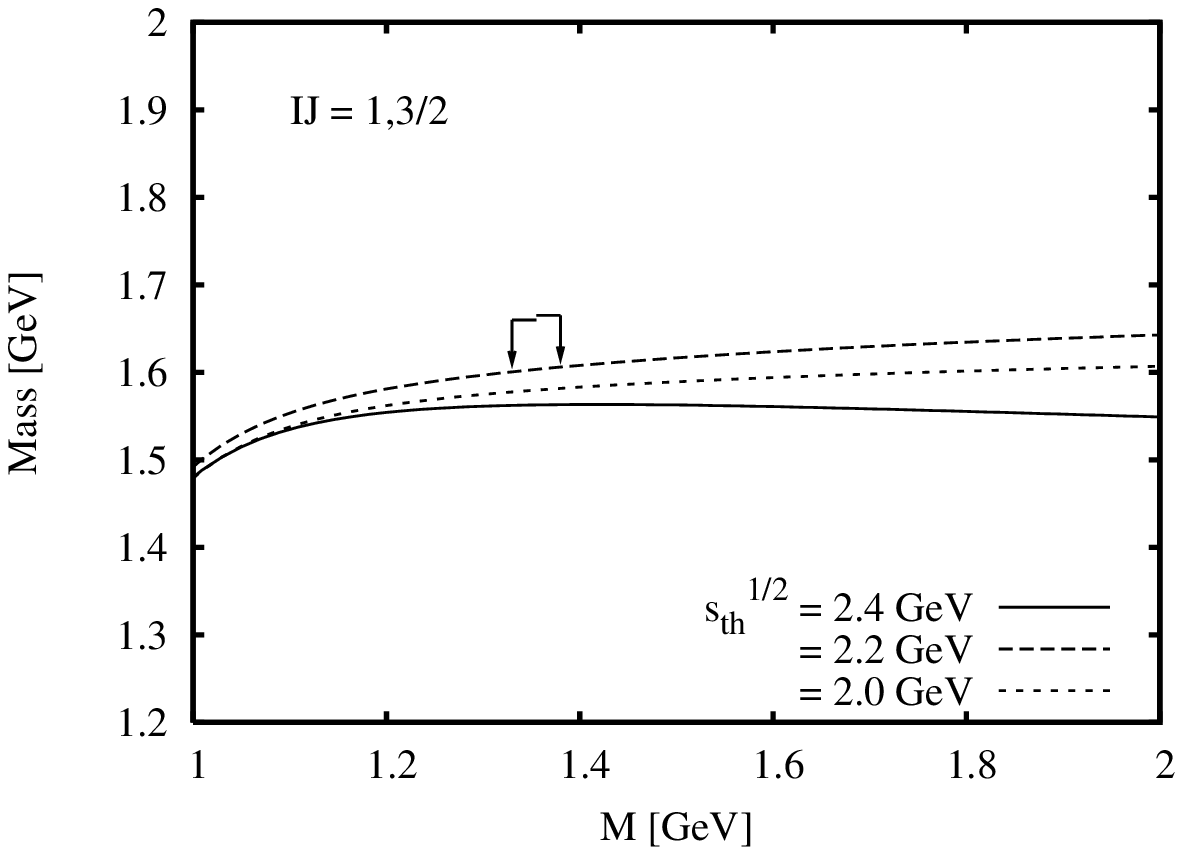} 
\caption{The mass of the pentaquark with quantum numbers $IJ = 0\frac{1}{2}$, 
$1\frac{1}{2}$, $0\frac{3}{2}$, $1\frac{3}{2}$ 
as a function of the Borel mass $M$, obtained from the sum rule of the 
chiral-even part. The arrows indicate the boundary of the Borel window 
for the middle value of the threshold parameter $s_{th}$.}
\label{fig:mass1}
\end{center}
\end{figure*}
We first show our obtained results for the isosinglet, spin $\frac{1}{2}$ case. This quantum number has been already 
frequently investigated as a possible assignment for $\Theta^+(1540)$ in QCD sum rules \cite{Kojo1,Zhu,Matheus1,Sugiyama,
Eidemuller,Ioffe,Kondo,Lee} and lattice QCD \cite{Sasaki,Ishii1,Takahashi,Lasscock1}. 
As for QCD sum rules, most of these calculations have problems in the establishment of
the Borel window. In \cite{Kojo1}, this problem was avoided
by taking the similar approach with that of the present paper,
but we found some mistakes in the computation of the OPE
beyond dimension 8. After correcting them,
we got the present result which excludes the positive
parity state obtained in \cite{Kojo1}.

\subsubsection{Sum rule for the chiral-even part}

Using the operators of Eqs.(\ref{eq:field1}),(\ref{eq:field2}), 
we first determine the mixing angle $\phi^0_{1/2}$ and the threshold parameter $s_{th}$. 
The values that we have obtained are, $\phi^0_{1/2} = -0.22$ and $\sqrt{s_{th}} = 2.2\,\mathrm{GeV}$. Furthermore, checking the convergence 
of the OPE and investigating the value of the pole contribution, we have confirmed that a Borel window exists for 
$1.2 \,\mathrm{GeV} \lesssim M \lesssim 1.6 \,\mathrm{GeV}$ (for details, consult Figs. \ref{fig:conv1}, \ref{fig:ope1} and 
\ref{fig:pole1} of Appendix \ref{apB}).

The calculated value of the ground state mass $m_{\Theta^+}(M, s_{th})$ of Eq.(\ref{eq:mass}) is given in 
Fig. \ref{fig:mass1} (top left) as a function of the 
Borel mass $M$. The boundary of the Borel window for the case of $\sqrt{s_{th}} = 2.2\,\mathrm{GeV}$ are indicated by the two arrows. 
One can see that the obtained value is about $1.5\,\mathrm{GeV}$ within the Borel window. Even though 
we have found a wide Borel window, the curves shown in Fig. \ref{fig:mass1} exhibit quite a large dependence on $M$ and $s_{th}$, which 
suggests that the spectral function only contains $KN$ scattering states and not a narrow pole. On the other hand, as will be shown later, 
the result of the parity-projected sum rules are fairly stable against $M$ and $s_{th}$, which rather points to a narrow  pole in the ground state. 
The interpretation these different results will be discussed below. 

\begin{figure*}
\begin{center}
\includegraphics[width=8cm,clip]{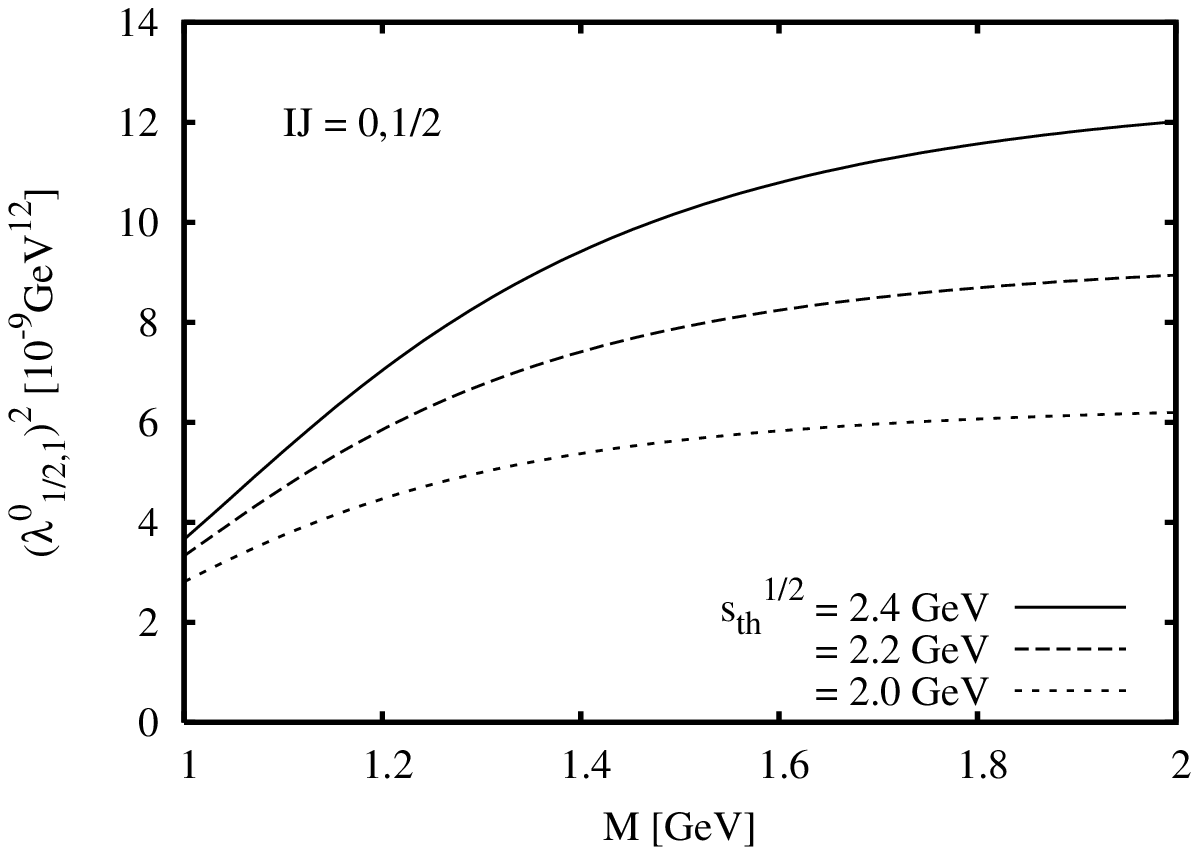}
\includegraphics[width=8cm,clip]{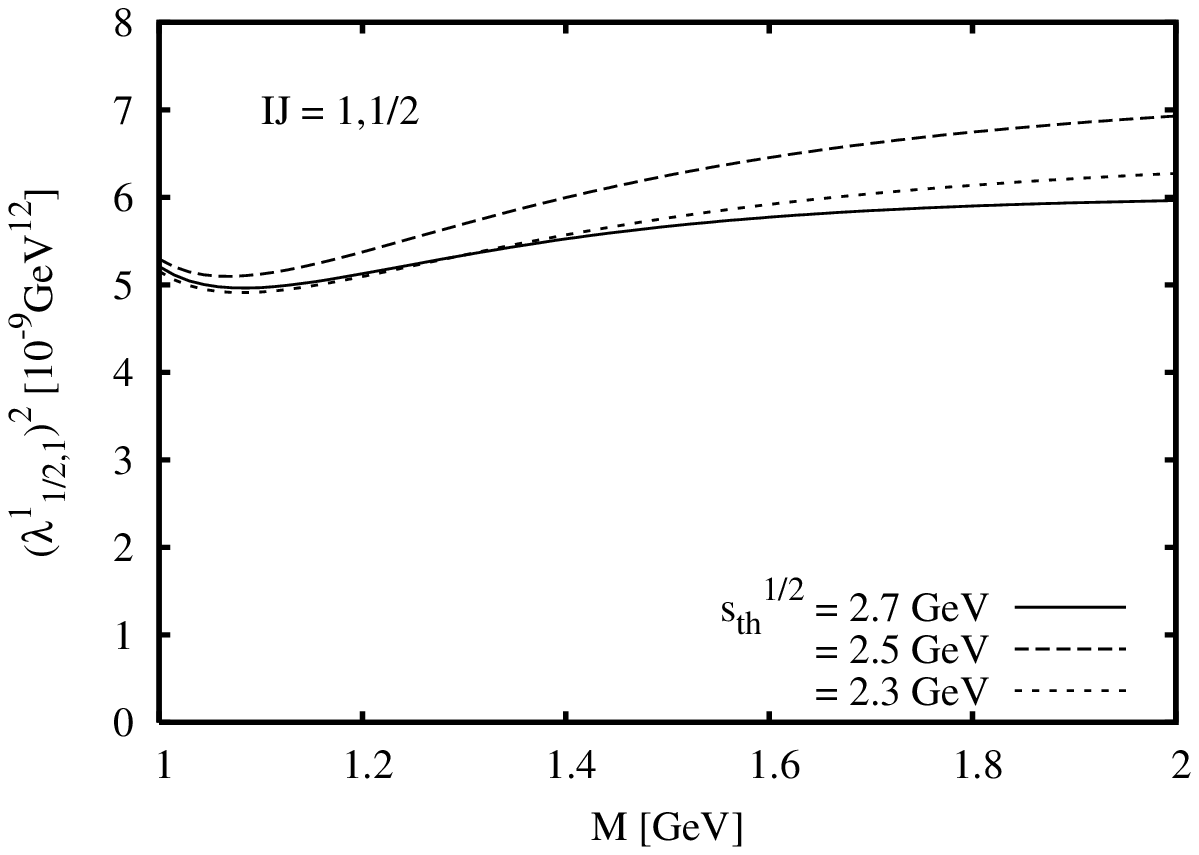}
\includegraphics[width=8cm,clip]{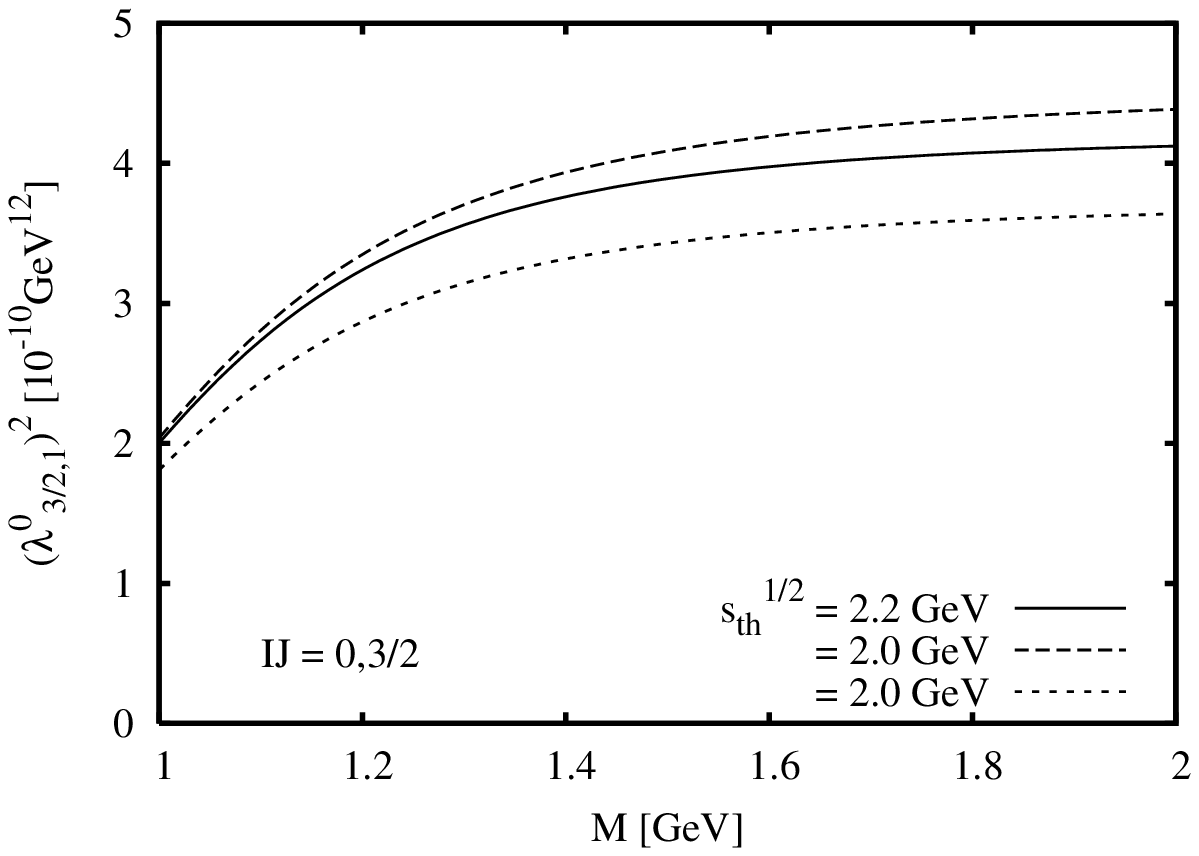}
\includegraphics[width=8cm,clip]{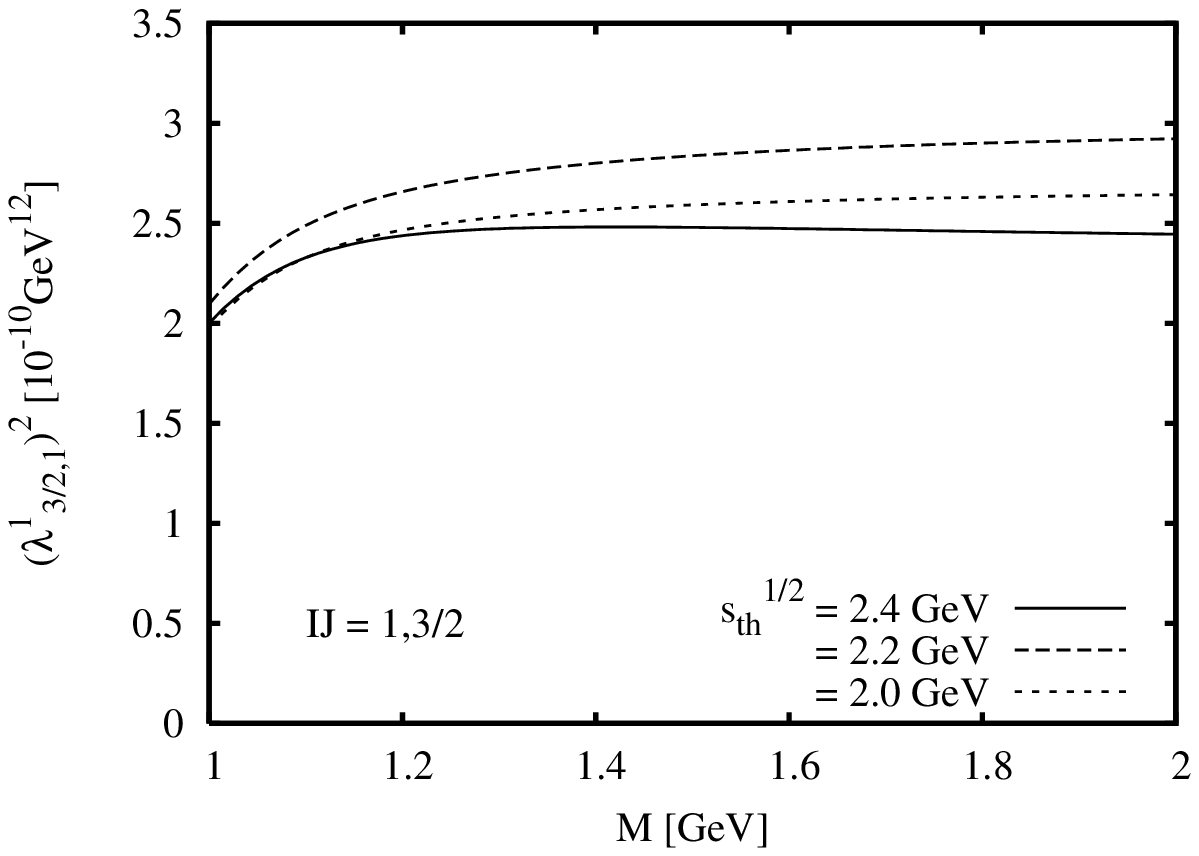}
\caption{The residue $(\lambda^I_{J,1})^2$ for the quantum numbers $IJ = 0\frac{1}{2}$, 
$1\frac{1}{2}$, $0\frac{3}{2}$, $1\frac{3}{2}$, 
obtained from Eq.(\ref{eq:residue}) for the sum of the chiral-even part. The value is given for 
three different threshold parameters.}
\label{fig:residue1}
\end{center}
\end{figure*}
The result of the residue, calculated from Eq.(\ref{eq:residue}) is given in Fig. \ref{fig:residue1}. As well as for the mass, 
the results for the residue depend on $M$ and $s_{th}$ quite strongly.

\subsubsection{Parity-projected sum rules}
\begin{figure*}
\begin{center}
\includegraphics[width=8cm,clip]{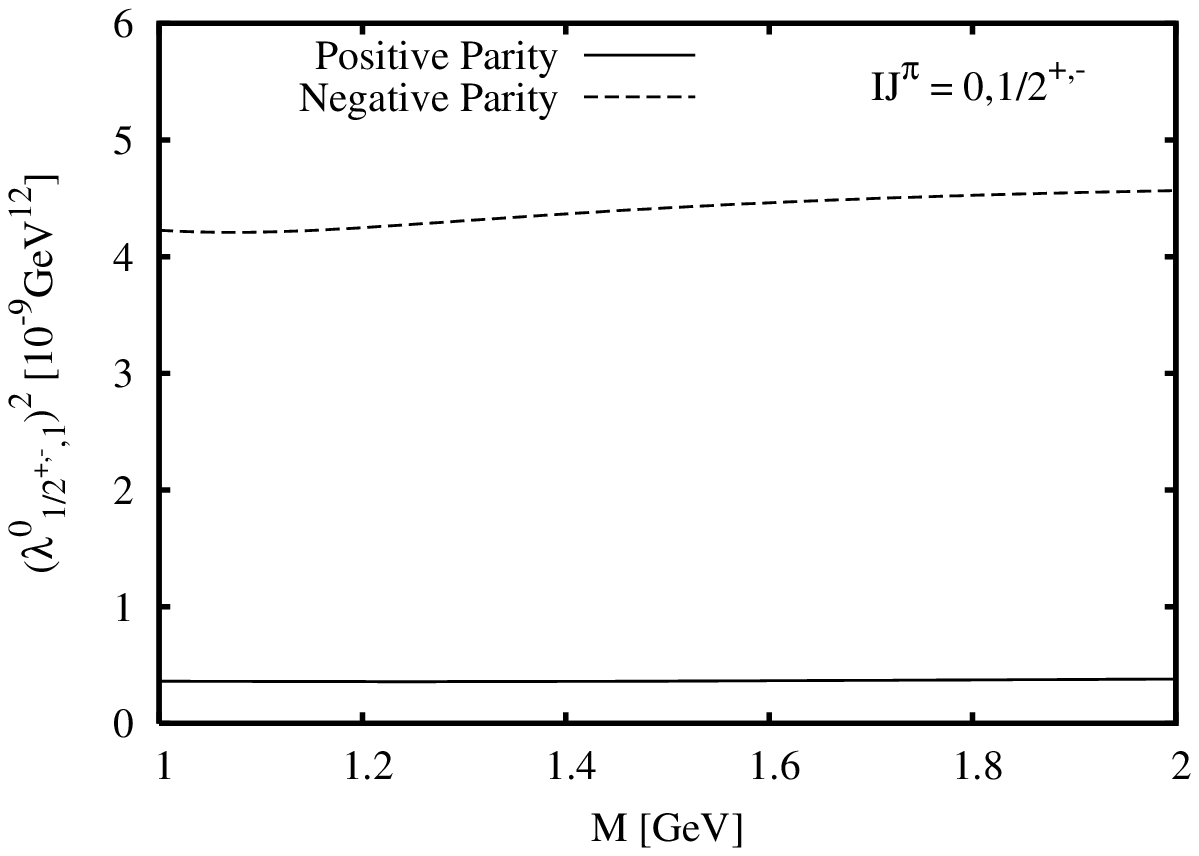}
\includegraphics[width=8cm,clip]{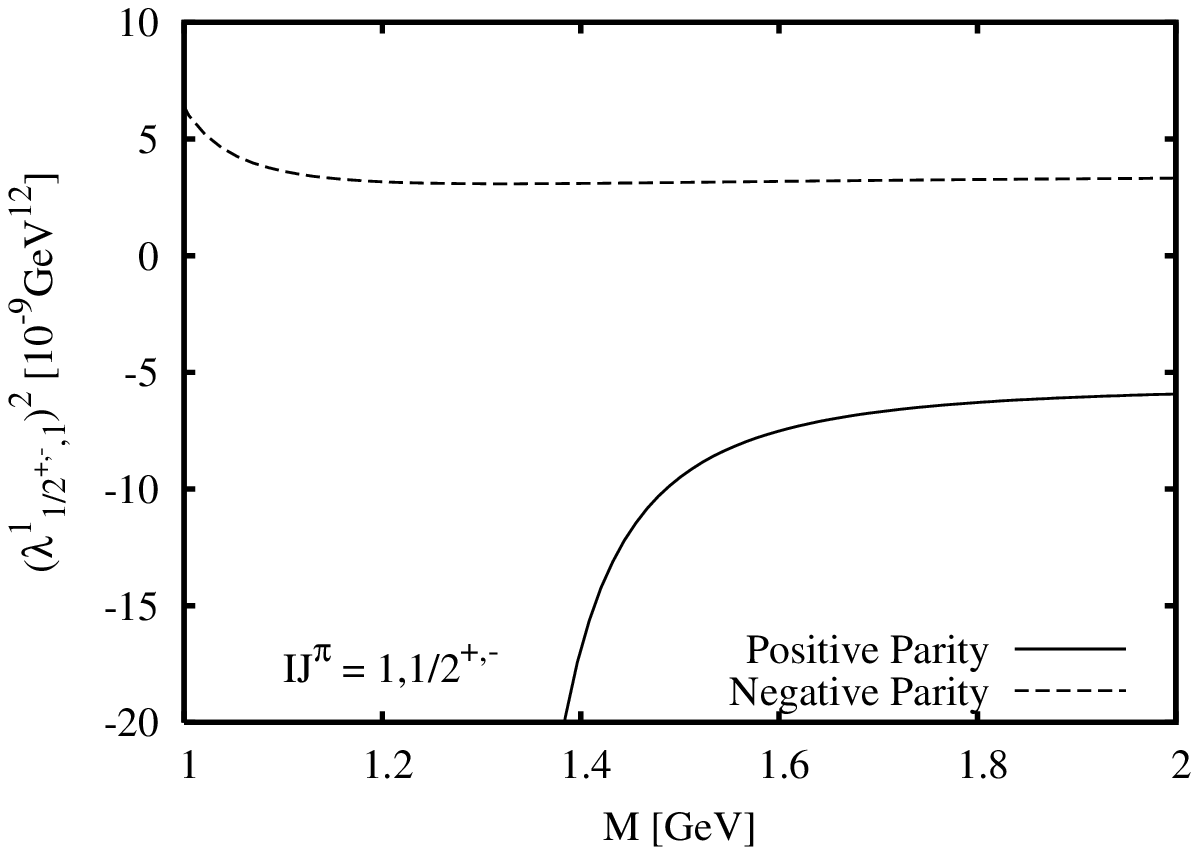}
\includegraphics[width=8cm,clip]{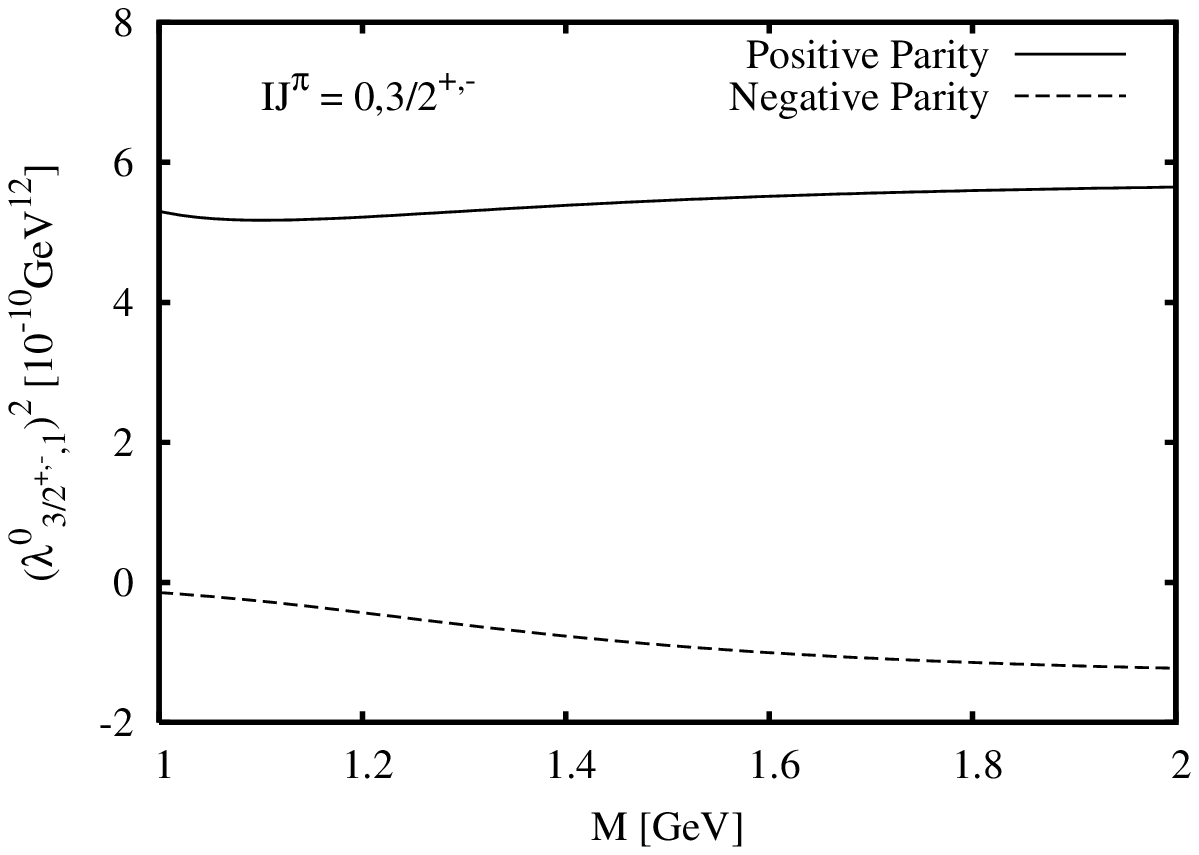}
\includegraphics[width=8cm,clip]{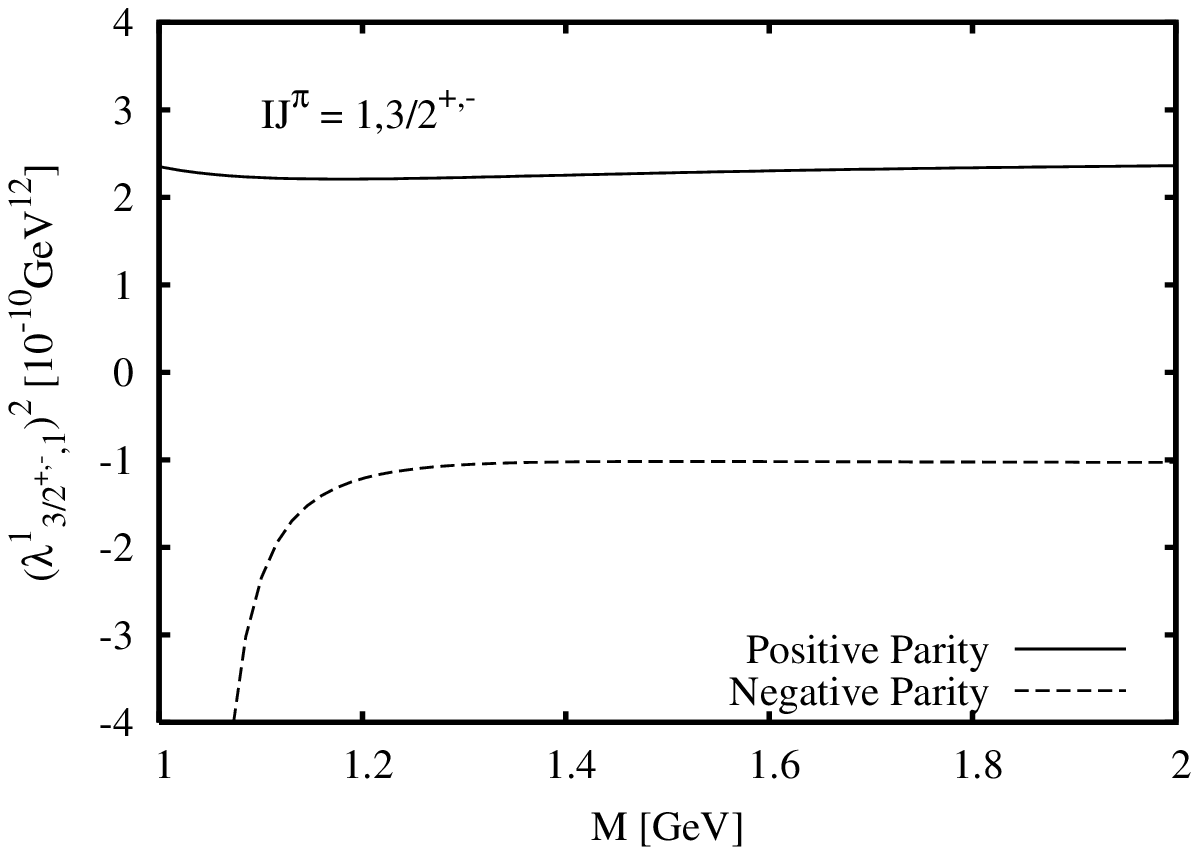}
\caption{The residues of the positive and negative parity sum rules for the quantum numbers $IJ = 0\frac{1}{2}$,  
$1\frac{1}{2}$, $0\frac{3}{2}$, $1\frac{3}{2}$, 
obtained from Eq.(\ref{eq:ope3}). The curves are calculated in the chiral limit ($m_s=0$, $\langle\overline{s}s\rangle = \langle\overline{q}q\rangle$).}
\label{fig:residue2}
\end{center}
\end{figure*}

\begin{figure*}
\begin{center}
\includegraphics[width=8cm,clip]{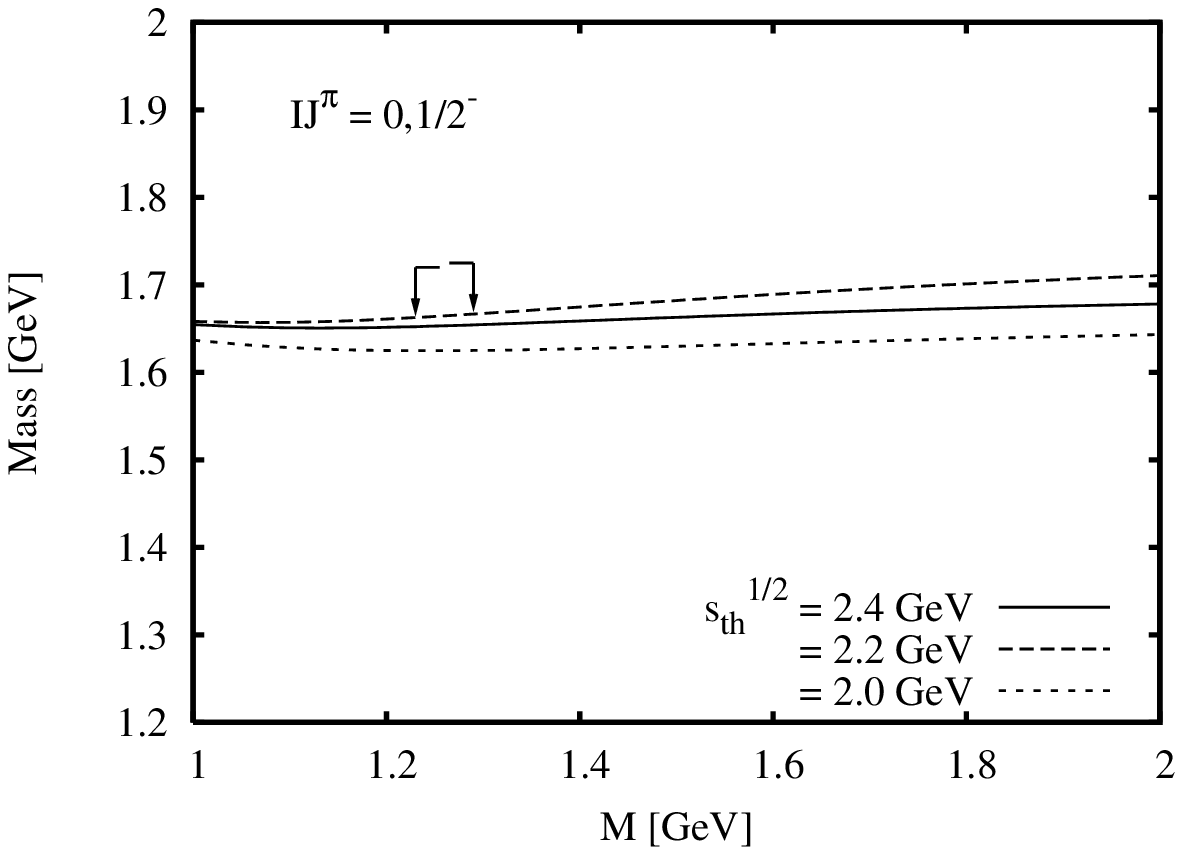}
\includegraphics[width=8cm,clip]{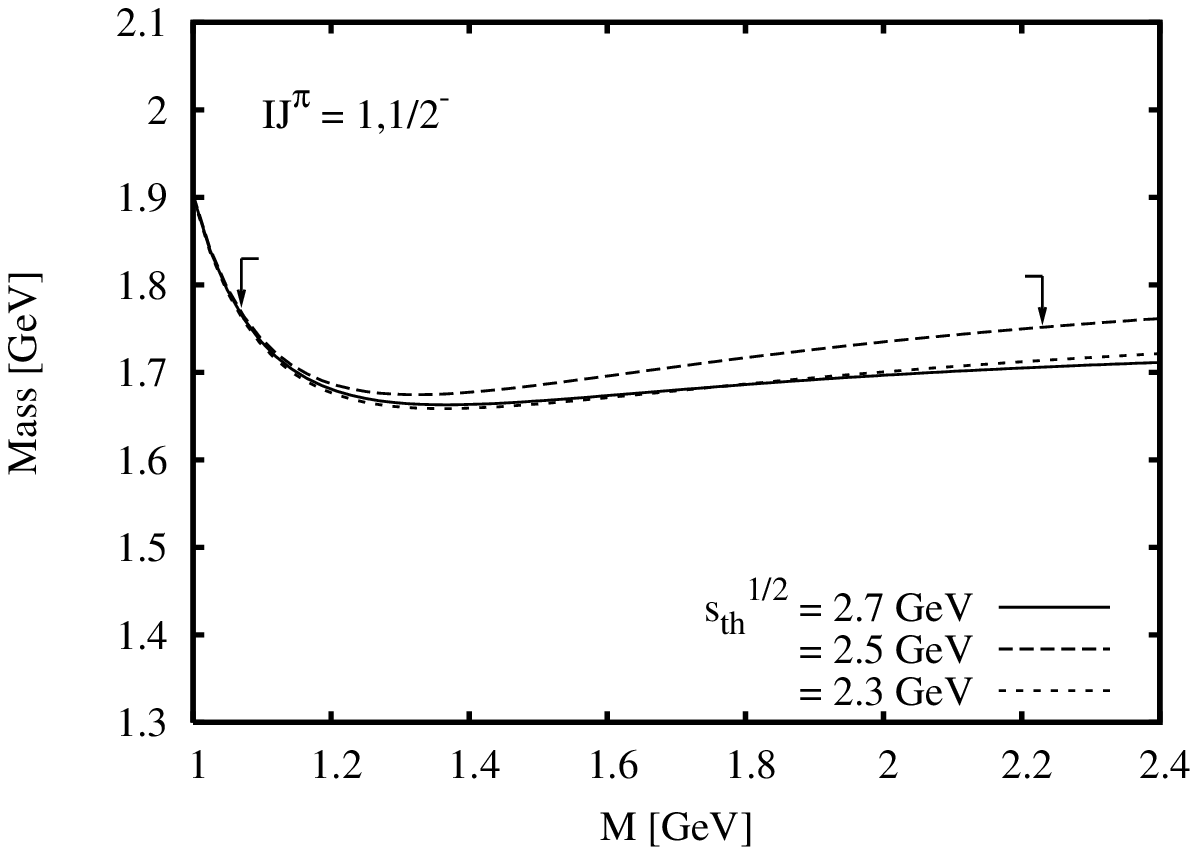}
\includegraphics[width=8cm,clip]{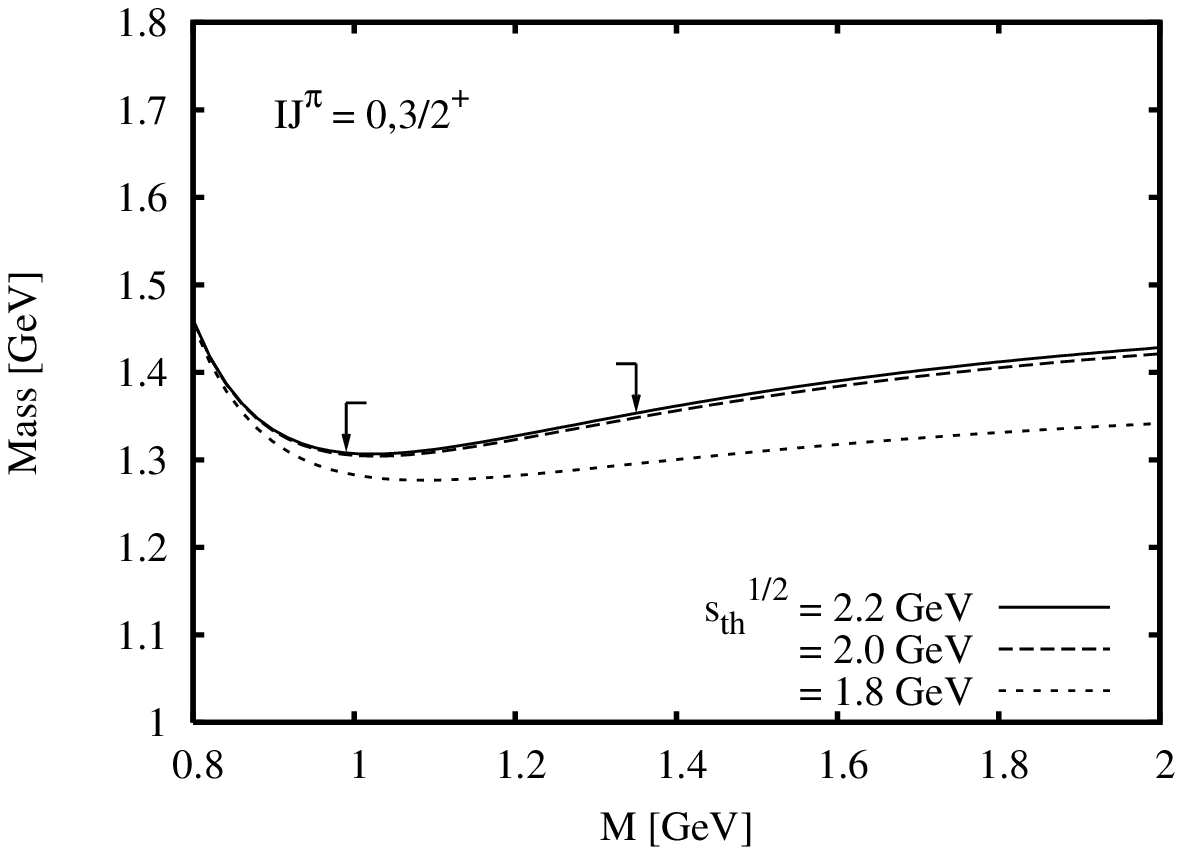}
\includegraphics[width=8cm,clip]{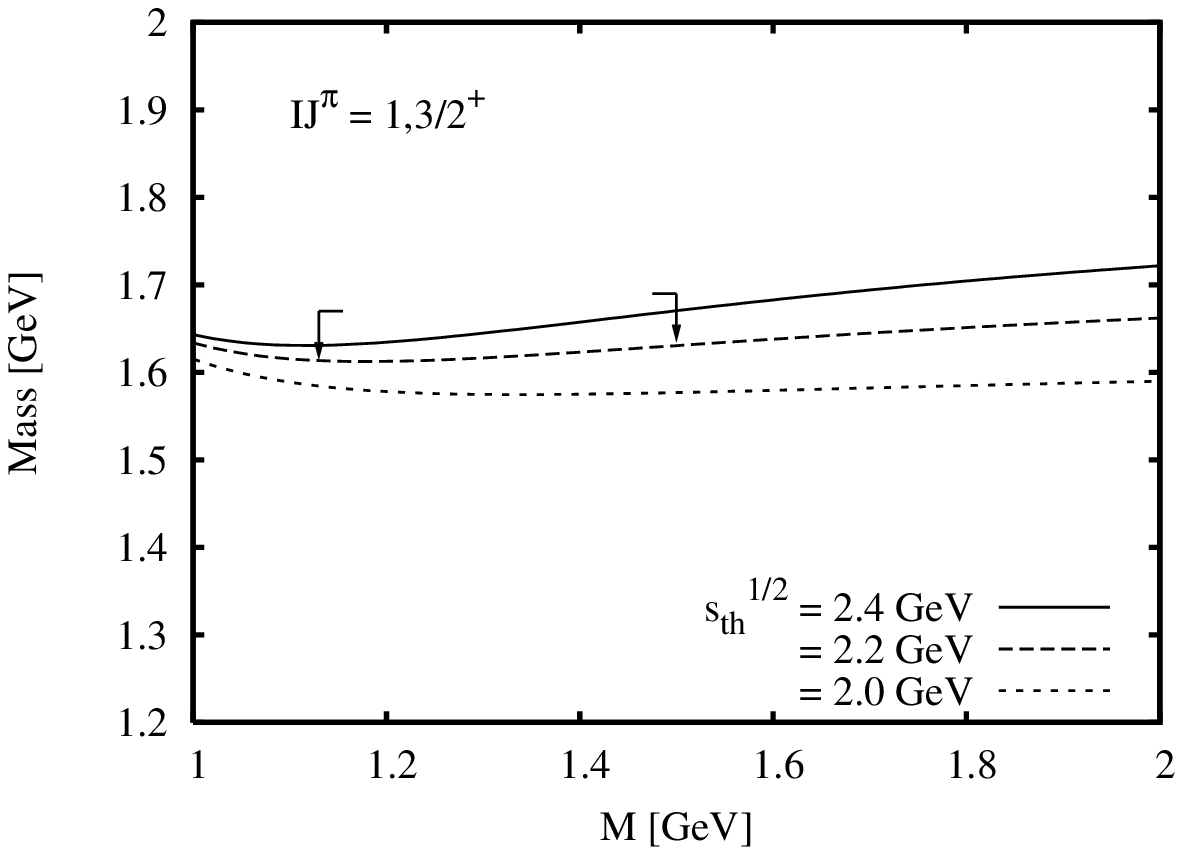}
\caption{The mass of the pentaquark for $IJ^{\pi} = 0\frac{1}{2}^-$, 
$IJ^{\pi} = 1\frac{1}{2}^-$, $IJ^{\pi} = 0\frac{3}{2}^+$, $IJ^{\pi} = 1\frac{3}{2}^+$ 
as a function of the Borel mass $M$. The arrows indicate 
the boundary of the Borel window for the middle value of the 
threshold parameter $s_{th}$. 
The curves are calculated in the chiral limit.}
\label{fig:mass2}
\end{center}
\end{figure*}

As already mentioned before, we will use the parity-projected sum rules in the chiral limit. This is justified, as we have confirmed in the sum 
rules of the chiral-even part that the qualitative behavior of the results does not change when this limit is taken. 
To show the strength of the contribution of the positive and negative parity states in the spectral function of the sum rule, 
the residues of the parity-projected sum rules [Eq.(\ref{eq:ope3})] are given in Fig. \ref{fig:residue2}.
It is clear from this figure that the negative parity states dominate and that therefore negative parity has to be assigned 
the state investigated in the last section. Furthermore, the mass calculated from the negative parity sum rule of Eq.(\ref{eq:ope3}) is 
shown in Fig. \ref{fig:mass2}, together with the Borel window for the threshold parameter $\sqrt{s_{th}} = 2.2\,\mathrm{GeV}$.
\begin{figure}
\includegraphics[width=8cm,clip]{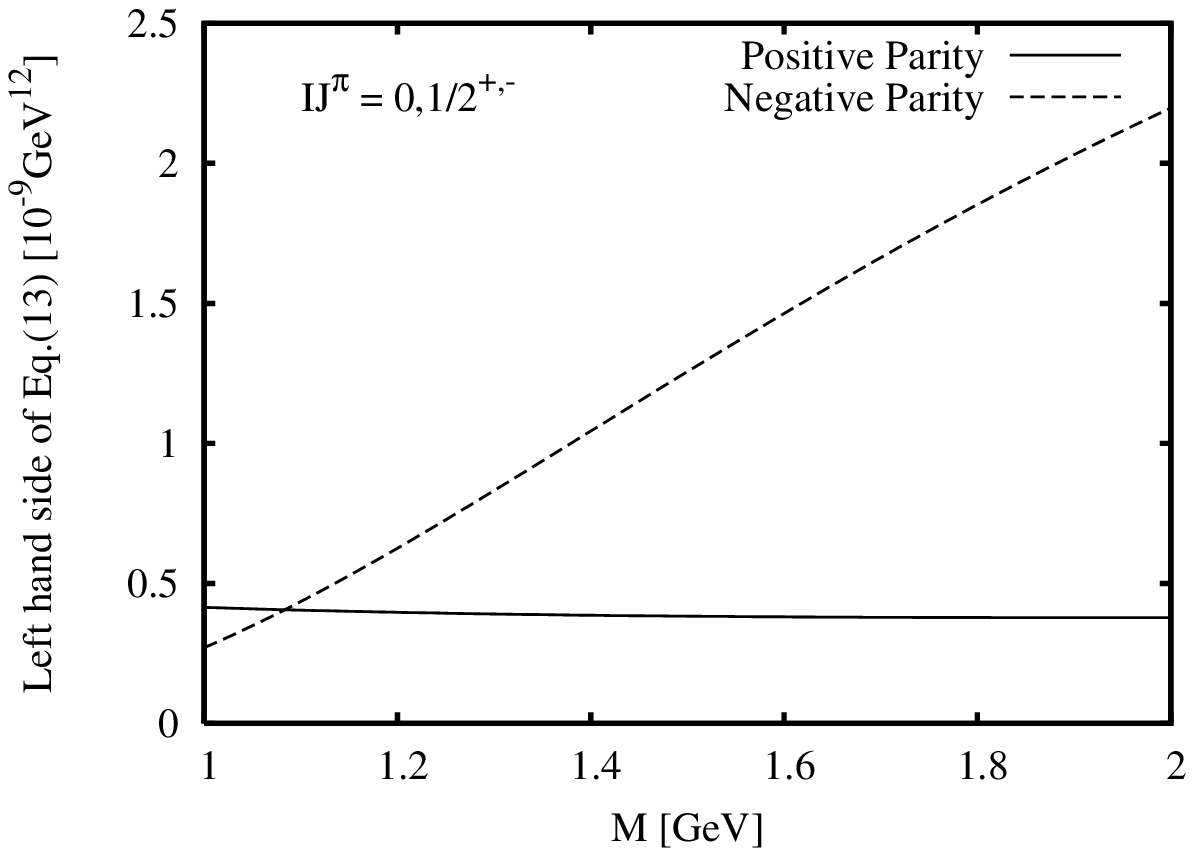}
\caption{The contribution of positive and negative parts of Eq.(\ref{eq:ope1}) for $IJ = 0\frac{1}{2}$, obtained from 
Eq.(\ref{eq:ope3}) in the chiral limit. The value of the threshold parameter is $\sqrt{s_{th}} = 2.2\,\mathrm{GeV}$.}
\label{fig:compare1}
\end{figure}
As is seen in the figure, a valid Borel window is established around $1.2 \,\mathrm{GeV} \lesssim M \lesssim 1.3 \,\mathrm{GeV}$ 
and the obtained value is consistent with the one of the chiral-even sum rule. Moreover, the dependencies on both $M$ and $s_{th}$ 
are very small, which in contrast to the chiral-even case rather points to a narrow ground state pole and not to 
$KN$ scattering states. 

It is puzzling why the behavior of these two sum rules is so different, even though the contribution of the positive parity states 
is very small, as shown in Fig. \ref{fig:residue2}. Numerically, this can be understood from the fact that the chiral-even part is 
multiplied by an additional power of $q_0$ in the parity-projected sum rules [compare Eqs.(\ref{eq:cor1}) and (\ref{eq:retard})], which 
considerably changes the behavior of the sum rules in this case. Moreover, we have confirmed that even though the residue for 
positive parity state is small, it numerically has a large influence on Eq.(\ref{eq:ope1}) for the low Borel mass region. 
To illustrate this point, the contribution of positive and and negative parts, calculated from 
Eq.(\ref{eq:ope3}) in the chiral limit, are shown in Fig. \ref{fig:compare1}. The negative parity part clearly shows an 
unphysical behavior as it is almost constant, while it should be an exponentially increasing function in the case of a narrow 
ground state pole dominating the sum rules. 
Nevertheless, around $1.2\,\mathrm{GeV}$, its contribution is comparable to positive parity part and therefore has 
a strong influence on the result of the chiral-even part. 
Thus, the most 
reasonable explanation for these different results seems to be that the positive parity $KN$ scattering states are contaminating 
the results of the chiral-even part and therefore lead to a large dependence on $M$ and especially on $s_{th}$. We hence 
conclude that we have found some real evidence for a narrow ground state pole with $IJ^{\pi}=0\frac{1}{2}^-$ even though the 
situation is more ambiguous than in the other channels.

\subsection{$IJ^{\pi} = 1\frac{1}{2}^{\pm}$}
Next, the isotriplet, spin $\frac{1}{2}$ states are studied. As no isospin partners of $\Theta^+(1540)$ have so far been 
found experimentally, it is currently believed to be an isosinglet state, but this assignment is not conclusive yet. Furthermore, 
even if $\Theta^+(1540)$ is an isosinglet state, a different isotriplet pentaquark state could exist at higher energies. We thus 
consider this state in the following paragraphs.  

\subsubsection{Sum rule for the chiral-even part}
The method is essentially parallel to the isosinglet case, the difference being only 
that we employ the operators of Eqs.(\ref{eq:field3}) and (\ref{eq:field4}) instead 
of Eqs.(\ref{eq:field1}) and (\ref{eq:field2}). 
The values of 
the mixing angle $\phi^{1}_{1/2}$ and the threshold parameter $s_{th}$ have turned out to be 
$\phi^{1}_{1/2} = -0.079$ and $\sqrt{s_{th}} = 2.5\,\mathrm{GeV}$.

The mass calculated from Eq.(\ref{eq:mass}) is shown in Fig. \ref{fig:mass1} (top right), as before with the Borel window for the middle value of the 
threshold parameter, indicated by the two arrows. The obtained value is about $1.6\,\mathrm{GeV}$. 
Compared to the isosinglet spin $\frac{1}{2}$ case of Fig. \ref{fig:mass1}, it is obvious that the dependence on the Borel mass $M$ and especially 
on the threshold parameter $s_{th}$ is small, which is positive evidence for a narrow ground state pole in the spectral function. 
The residues for the three different threshold parameters are given in 
Fig. \ref{fig:residue1}, where 
we again get only a similarly mild dependence on $M$ and $s_{th}$. 

\subsubsection{Parity-projected sum rules}
We will follow the same method as in the isosinglet case and calculate the parity-projected sum rules in the 
chiral limit. As is shown in Table \ref{dependencies}, the result of the chiral-even sum rule did depend on the strange quark 
mass $m_s$ quite strongly and one thus may wonder whether the procedure of taking the chiral limit is justified. But, as 
we will use this sum rule only to determine the parity of the state, we think that it is accurate enough to provide reliable 
information, because even though the mass value of the state may quantitatively change, it is improbable that the parity 
of the state will switch when this limit is taken.

The residues of the positive and negative parity sum rules are compared in Fig. \ref{fig:residue2}.
In this figure, it is seen that both residues are similar in magnitude. (Note that, as we have taken the difference of two 
correlators, the residue $(\lambda^1_{1/2,1})^2$ can become negative. States with negative residues can thus not 
be ruled out as unphysical like in the ordinary QCD sum rules with just one correlator.) On the other hand, the residue of 
the positive parity state is very unstable against the variation of the Borel mass, which suggests that it does not 
correspond to a narrow 
ground state pole that we are looking for. Meanwhile, the residue with negative parity is fairly stable and thus seems to be 
consistent with the chiral-even sum rule.

The masses of both parity states are also calculated. As expected form the result of the residues, the calculated mass of 
positive parity strongly depends on the Borel mass $M$ and no stable region is found. In contrast, the results for negative parity 
are stable and consistent with the value obtained from the chiral-even sum rule. We therefore conclude that the parity of the 
state is negative. The mass values for the negative parity case are shown in Fig. \ref{fig:mass2}.

Compared to all other cases studied in this paper, the Borel window here 
seems to be unnaturally large. The reason for this is that the same phenomenon as in the upper left part of 
Fig. \ref{fig:pole1} has occurred, 
meaning that due to some cancellation in the integral of the spectral function above $s_{th}$, a peak has emerged in 
the function of the pole contribution, which shifts the upper boundary of the Borel window to a high value and 
therefore leads to this very large Borel window. 

\subsection{$IJ^{\pi} = 0\frac{3}{2}^{\pm}$}
This quantum number has been already investigated in detail by the present authors in a recent paper \cite{Gubler}. 
We will not repeat the analysis given there and only restate the most important results.

The same strategy as in this paper 
was followed, meaning that the difference of two correlators was taken, and the values of the mixing 
angle and threshold parameter were determined from the conditions of pole dominance. The obtained values are 
$\phi^0_{3/2} = 0.063$ and $\sqrt{s_{th}} = 2.0\,\mathrm{GeV}$. This then leads to the mass values 
shown in Fig. \ref{fig:mass1} (bottom left), calculated from the chiral-even sum rule. The obtained value lies at about 
$1.4\,\mathrm{GeV}$.
The result shows both a small dependence on $M$ and $s_{th}$, which suggests that a narrow 
ground state pole exists in the spectral function of this quantum number. 

The parity of the state is determined with the parity-projected sum rules, leading to 
Figs. \ref{fig:residue2} and \ref{fig:mass2}. Fig. \ref{fig:residue2} shows that the pole strength is 
dominated by the residue of the positive parity state. Fig. \ref{fig:mass2} then confirms that the 
positive parity sum rules give stable results, which are consistent with the ones obtained from the 
chiral-even sum rule.

\subsection{$IJ^{\pi} = 1\frac{3}{2}^{\pm}$}
The existence of
states with quantum numbers $IJ^P = 1\frac{3}{2}^{\pm}$ have been suggested for instance 
by studies using the quark model \cite{Kanada} and the chiral 
unitary approach \cite{Sarkar}. We further investigate them here using the QCD sum rule method.

\subsubsection{Sum rule for the chiral-even part}
The operators used are given in 
Eqs. (\ref{eq:field7}) and (\ref{eq:field8}) and the following values have been obtained 
for the mixing angle and the threshold parameter:
$\phi^1_{3/2} = 0.024$ and $\sqrt{s_{th}} = 2.2\,\mathrm{GeV}$.

The results for the mass are shown in Fig. \ref{fig:mass1} (bottom right) together with 
the Borel window for $\sqrt{s_{th}} = 2.2\,\mathrm{GeV}$. As can be read off 
from the figure, a value around $1.6\,\mathrm{GeV}$ is obtained for the mass of the state. 
The dependence of the result on both $M$ and $s_{th}$ is weak, which suggests that a narrow pole is present in the spectrum. 

The value of the residue $(\lambda^{1}_{3/2,1})^2$ is given in Fig. \ref{fig:residue1}, 
where again only a small dependence on $M$ and $s_{th}$ is observed.

\subsubsection{Parity-projected sum rules}
We have obtained 
a consistent result for the positive parity channel, while no state below $2.0\,\mathrm{GeV}$ was found with negative parity. 
The two residues are shown in Fig. \ref{fig:residue2}, where one can see that the magnitude of the positive parity residue is 
larger that the one of negative parity and that it is an almost completely stable against the variation of $M$. 
The calculated mass of the positive parity sum rule, shown in Fig. \ref{fig:mass2}, moreover gives 
similar values as obtained in the chiral-even case, which do not strongly depend on $M$ and $s_{th}$. 
We therefore conclude that positive parity has to be assigned to the investigated state.
\begin{table*}
\begin{center}
\caption{Contributions of the uncertainties of all parameters appearing in the calculation to 
the final error. Only values larger than $\pm0.05\,\mathrm{GeV}$ 
are explicitly given. These values have been obtained using the sum rule of the chiral-even part.} 
\label{dependencies}
\begin{tabular}{lcccc} \hline
$IJ^P=$ \hspace*{1cm}& \hspace*{1cm} $0\frac{1}{2}^{\pm}$\hspace*{1cm}& \hspace*{1cm}$1\frac{1}{2}^{\pm}$\hspace*{1cm}& 
\hspace*{1cm}$0\frac{3}{2}^{\pm}$\hspace*{1cm}& \hspace*{1cm} $1\frac{3}{2}^{\pm}$ \hspace*{1cm} \\ \hline
$M$ & $\pm0.10\,\,\mathrm{GeV}$ & $\pm0.05\,\,\mathrm{GeV}$ & $\sim 0$ & $\sim 0$ \\
$s_{th}$ & $\pm0.10\,\,\mathrm{GeV}$ & $\sim 0$ & $\pm0.05\,\,\mathrm{GeV}$ & $\pm0.05\,\,\mathrm{GeV}$\\
$\langle\overline{q}q\rangle$ & $\pm0.15\,\,\mathrm{GeV}$ & $\pm0.20\,\,\mathrm{GeV}$ & $\sim 0$ & $\pm0.10\,\,\mathrm{GeV}$\\
$\frac{\langle\overline{s}s\rangle}{\langle\overline{q}q\rangle}$ & $\sim 0$ & $\pm0.05\,\,\mathrm{GeV}$ & $\sim 0$ & $\sim 0$\\
$\frac{\langle\overline{q}g \sigma\cdot G q\rangle}{\langle\overline{q} q \rangle}$ & $\pm0.05 \,\,\mathrm{GeV}$ & $\sim 0$ & 
$\pm0.10\,\,\mathrm{GeV}$ & $\pm0.05\,\,\mathrm{GeV}$\\
$\frac{\langle\overline{s}g \sigma\cdot G s\rangle}{\langle\overline{s} s \rangle}$ & $\sim 0$ & $\sim 0$ & $\sim 0$ & $\sim 0$\\
$\langle\frac{\alpha_s}{\pi}G^2\rangle$ & $\pm0.05\,\,\mathrm{GeV}$ & $\pm0.20\,\,\mathrm{GeV}$ & $\sim 0$ & $\sim 0$\\ 
$m_s$ & $\pm0.05\,\,\mathrm{GeV}$ & $\pm0.20\,\,\mathrm{GeV}$ & $\sim 0$ & $\pm0.05\,\,\mathrm{GeV}$\\
$\kappa$ & $+0.15\,\,\mathrm{GeV}$ & $+0.20\,\,\mathrm{GeV}$ & $+0.05\,\,\mathrm{GeV}$ & $+0.10\,\,\mathrm{GeV}$\\ \hline
combined error & $\pm0.3\,\,\mathrm{GeV}$ & $\pm0.4\,\,\mathrm{GeV}$ & $\pm0.2\,\,\mathrm{GeV}$ & $\pm0.3\,\,\mathrm{GeV}$ \\ \hline
\end{tabular}
\end{center}
\end{table*} 

\subsection{Estimation of the theoretical ambiguity}
As the last point, we have to investigate the dependencies of the results on the various parameters of 
Table \ref{parameters}, in order to 
obtain a quantitative estimate of the error inherent in our results. 
We will here use only the results of the chiral-even part for this estimation. 

The contributions to the errors for the different quantum numbers are given in Table \ref{dependencies}.  
For instance, considering the $IJ^P = 0\frac{1}{2}^{\pm}$ case, 
we have already seen from Fig. \ref{fig:mass1}, that the 
dependence of the mass value on $M$ or $s_{th}$ leads to an uncertainty of about $\pm0.1\,\mathrm{GeV}$.
Among the other parameters, the result depends most strongly 
on $\langle\overline{q}q\rangle$, which gives an uncertainty 
of about $\pm0.15\,\mathrm{GeV}$. 
Similarly, raising the breaking parameter of the vacuum saturation approximation to $\kappa = 2$ leads to an 
increase of the mass of about $0.15\,\mathrm{GeV}$. Similar considerations lead to all the 
error contributions for the other quantum numbers given in Table \ref{dependencies}. 

Assuming that the various errors are uncorrelated, 
the final error estimations are then obtained by 
taking the root of the sum of all squared errors $\delta m_i$ and rounding up:
\begin{equation}
\text{combined error} \sim \sqrt{\sum_i (\delta m_i)^2}.
\end{equation} 
Note that this is merely a rough estimation, as there are additional errors 
coming from the truncation of the OPE and possible radiative corrections, 
that have been neglected in the current calculation. 
\begin{table*}
\begin{center}
\caption{Summarized results for all quantum numbers that have been investigated.
The allowed $KN$ decay channels of the respective quantum numbers are indicated in brackets. 
The mass values quoted here are obtained for the sum rule of the chiral-even part.} 
\label{results}
\begin{tabular}{|c|c|c|c|} \hline
  \multicolumn{1}{|c}{} & \multicolumn{1}{c|}{}&
  \multicolumn{2}{c|}{Parity} \\ \cline{3-4}
  \multicolumn{1}{|c}{} & \multicolumn{1}{c|}{}&
  \multicolumn{1}{c|}{+} &
  \multicolumn{1}{c|}{-} \\ \hline
  $\,J=\frac{1}{2}\,$    &  $\,I=0\,$  &\multicolumn{1}{c|}{\,no state found below $2.0$ GeV\,} & \multicolumn{1}{c|} {$1.5\pm0.3$ GeV (?)} \\ 
   & & \multicolumn{1}{c|}{\footnotesize{($KN$ P-wave)}} &\multicolumn{1}{c|}{ \footnotesize{($KN$ S-wave)}} \\ \cline{2-4}
   &  $I=1$  &\multicolumn{1}{c|} {no state found below $2.0$ GeV} &\multicolumn{1}{c|} {$1.6\pm0.4$ GeV} \\ 
   & & \multicolumn{1}{c|}{\footnotesize{($KN$ P-wave)}} &\multicolumn{1}{c|}{ \footnotesize{($KN$ S-wave)}} \\ \hline
  $J=\frac{3}{2}$   &  $I=0$  &\multicolumn{1}{c|} {$1.4\pm0.2$ GeV} &\multicolumn{1}{c|} {\,no state found below $2.0$ GeV\,} \\ 
   & & \multicolumn{1}{c|}{\footnotesize{($KN$ P-wave)}} &\multicolumn{1}{c|}{ \footnotesize{($KN$ D-wave)}} \\ \cline{2-4}
   &  $I=1$  &\multicolumn{1}{c|} {$1.6\pm0.3$ GeV}&\multicolumn{1}{c|}{no state found below $2.0$ GeV} \\ 
   & & \multicolumn{1}{c|}{\footnotesize{($KN$ P-wave)}} &\multicolumn{1}{c|}{ \footnotesize{($KN$ D-wave)}} \\ \hline
\end{tabular}
\end{center}
\end{table*}

\section{Discussion}
The details of the results for the various quantum numbers have been presented in the last section. 
Putting everything together, these results can be summarized as in Table \ref{results}. 

A number of comments have to be made here. First of all, the statement ``no state found below $2.0$ GeV" 
in Table \ref{results} means 
that either no valid Borel window could be found or that the results of the sum rules did strongly depend on $M$ and $s_{th}$ and that 
therefore no evidence for a narrow ground state pole could be found. Concerning this point, 
in the case of $IJ^{\pi} = 0\frac{1}{2}^{-}$, the results of 
the chiral-even sum rule and the parity-projected sum rule are to a certain extent contradictory and we therefore have to put a 
question mark behind this conclusion. Furthermore, having found no narrow state in our sum rule calculation does not 
necessarily mean that such a state 
does not exist. It could happen that the spectral function is dominated by the $KN$ scattering states and that
the narrow states that we are looking for only couple weakly to the interpolating field that we have used. 
Nevertheless, we have constructed general operators from linear combinations of two independent local operators, and 
have investigated all possible mixing angles and therefore the nonobservation of states shown 
in Table \ref{results} has been confirmed for quite a wide range of independent operators.

Next, we discuss the physical implications of the obtained results. A question that comes to one's mind when looking 
at Table \ref{results} is, why we can not observe $J^{\pi} = \frac{1}{2}^{+}$ states while we are seeing the ones with 
$J^{\pi} = \frac{3}{2}^{+}$. These states are in some models considered to be spin-orbit partners \cite{Jaffe}, so if these models 
are realistic and consistent with QCD, we should be able to observe both of these states. There are at least two 
possible explanations for our obtained results. One explanation could be that the states with $J^{\pi} = \frac{1}{2}^{+}$ in 
fact exist, but their coupling to the used operators are too small and/or the $KN$ scattering contribution is 
too large, so that a narrow peak structure cannot be extracted. Another possible interpretation of the missing 
$J^{\pi} = \frac{1}{2}^{+}$ states could be that, the spin-orbit partners of the spin $\frac{3}{2}$ states are not the 
ones with spin $\frac{1}{2}$ but with spin $\frac{5}{2}$. This would mean that $\Theta^+(1540)$ is 
indeed a very exotic state, as in this case the $uudd$ quarks 
have to form a spin 2 state, which would then couple to the remaining $\overline{s}$. This is of course only a very speculative 
picture, but it would be interesting to test it by calculating pentaquark states with spin $\frac{5}{2}$.

Another important point, that needs to be discussed, is the interpretation of our results on the $J^{\pi} = \frac{1}{2}^{-}$ states. 
Such a state was also found in a lattice study (conducted only for the isosinglet state), where a resonance state was isolated from 
the $KN$ scattering states \cite{Takahashi}. Our results (especially in the isosinglet case) are somewhat ambiguous, and 
the errors are large, so it is difficult to draw any definite conclusions. In any case, whether such states turn out to 
be real pentaquark resonances or not, they most possibly do not correspond to the observed $\Theta^+(1540)$ state, because 
$J^{\pi} = \frac{1}{2}^{-}$ states can decay into $KN$ by an S-wave, for which the width is expected to be much larger than 
the observed value for $\Theta^+$, which is less than $1\,\mathrm{MeV}$ \cite{Barmin2}. Of course, in principle 
there may exist some 
so far unknown mechanism, which suppresses the width strongly and which would allow to assign the $J^{\pi} = \frac{1}{2}^{-}$ 
quantum numbers to the $\Theta^+$, but with our present knowledge and experience, this seems to be unlikely.

\section{Conclusion}
We conclude from our results summarized in Table \ref{results} 
that the most probable quantum number candidate for $\Theta^+(1540)$ is $IJ^{\pi} = 0\frac{3}{2}^{+}$. 
We have also found evidence for 
an isotriplet state $1\frac{3}{2}^{+}$ and 
two states with spin $\frac{1}{2}$ 
($0\frac{1}{2}^-$ and $1\frac{1}{2}^-$) at slightly higher energy. 

To obtain these results, we have employed the QCD sum rule method, whose reliability 
is improved by analyzing the difference of two independent correlators, by which the contribution of the high-energy 
continuum states is suppressed. Furthermore, by calculating the OPE up to dimension 14 it is made sure that the expansion 
is converging well, and a valid Borel window can be established. 

Considering the spin $\frac{1}{2}$ states, 
although we could observe some evidence for resonance states with $IJ^{\pi} = 0\frac{1}{2}^{-}$ and $1\frac{1}{2}^{-}$ in the 
region of $1.5\,\mathrm{GeV}$, we 
have pointed out that the situation concerning the $KN$ 
scattering states does not seem to be very clear and our predictive power is quantitatively very limited. 
Furthermore, as discussed in the previous section, we do not believe that these states correspond to 
$\Theta^+(1540)$, because their width is expected to be too large to be consistent with the experimental value. 

Looking at the states with $IJ^{\pi} = 0\frac{3}{2}^{+}$ and $1\frac{3}{2}^{+}$, in both cases 
the values of the masses and residues show only a weak dependence on the Borel mass $M$ and the threshold 
parameter $s_{th}$. 
For the isosinglet case this was 
already pointed out in \cite{Gubler}. 
This suggests that we are really observing narrow resonance states in the spectral functions of these quantum numbers. 
As no isospin partners of the $\Theta^+(1540)$ have so far been found, it is believed to be an isosinglet, which leads 
to our conclusion that the $\Theta^+(1540)$is likely to be a state with quantum numbers $IJ^{\pi} = 0\frac{3}{2}^{+}$.  
The isotriplet state $1\frac{3}{2}^{+}$ is predicted to exist somewhere above the isosinglet, so 
it may be interesting for future experiments to look for this state. 
One nevertheless has to be cautious when interpreting the 
current results, as we cannot 
make any real quantitative prediction about the width of the state with the present method. Therefore it is difficult 
to say whether the predicted isotriplet state 
is narrow enough to be unambiguously detected in an experiment. 

The width of a state can be obtained 
from the QCD sum rule technique by calculating three-point functions of appropriate currents, and it would be interesting 
to see whether 
it is possible to obtain a value consistent with experiment for the $0\frac{3}{2}^{+}$ state and 
whether the $IJ^{\pi} = 1\frac{3}{2}^{+}$ state is really narrow enough to be experimentally observed. 
Furthermore, it is important to 
check whether our conjecture of the large widths of the $J^{\pi} = \frac{1}{2}^{-}$ states is really true or not. 
These issues are left for further studies.

\begin{acknowledgments}
This work was partially supported by KAKENHI under Contract Nos. 17070002 (Priority area), 
19540275, and 20028004. 
A part of this work was done in the Yukawa International Project for
Quark-Hadron Sciences (YIPQS). 
P.G. acknowledges the support by the Japan Society for the 
Promotion of Science for Young Scientists and is thankful for the hospitality of the Yukawa Institute for 
Theoretical Physics at Kyoto University, 
where part of this work has been completed. 
T.K. is supported by RIKEN, Brookhaven National Laboratory and the U. S.
Department of Energy under Contract No. DE-AC02-98CH10886.
\end{acknowledgments}

\onecolumngrid
\appendix
\section{Results of the operator product expansion \label{apA}}
We obtain the following result for the OPE 
in terms of the parameters $C_i$ defined in Eq.(\ref{eq:ope}). 
Note, that we here give the values of $C_i$ after the difference of the two 
correlators is taken, and that we have 
used $\theta^{I}_{J}-\theta^{I'}_{J} = \frac{\pi}{2}$ and $\phi^I_{J} = \theta^{I}_{J}+\theta^{I'}_{J}$.
After showing the results of the chiral-even part (up to terms proportional to $m_s$), 
the chiral-odd part is given in the chiral limit.

The used abbreviations are $G^2 \equiv G^a_{\mu\nu}G^{a\mu\nu}$ and 
$\sigma \cdot G \equiv \sigma^{\mu\nu} \frac{\lambda^a}{2} G^a_{\mu\nu}$, the $\lambda^a$ being the Gell-Mann 
matrices. $g$ is the coupling constant of QCD, giving $\alpha_s = \frac{g^2}{4\pi}$. The values of the condensates 
and the strange quark mass are shown in Table \ref{parameters}.
\subsection{$IJ^P = 0\frac{1}{2}^{\pm}$}
\subsubsection{Chiral-even part}
\begin{align}
C_0 =& 0, \mspace{18mu} C_4 = -\frac{ \langle\frac{\alpha_s}{\pi}G^2\rangle }{2^{14}3\cdot 5\pi^6}  \cos\phi^0_{1/2}, \mspace{18mu} 
C_6 = -\frac{\langle\overline{q}q\rangle^2}{2^73^2\pi^4}\sin\phi^0_{1/2} 
- \frac{m_s\langle\overline{s}g\sigma \cdot G s\rangle}{2^{13}\pi^6} \cos\phi^0_{1/2}, \notag \\
C_8 =& \frac{\langle\overline{q}q\rangle\langle\overline{q}g\sigma \cdot G q\rangle}{2^{9}3^2 \pi^4}  (7\cos\phi^0_{1/2} +34\sin\phi^0_{1/2}), \notag \\
C_{10} =& -\frac{\langle\overline{q}g\sigma \cdot G q\rangle^2}{2^{13}3^2 \pi^4} (22\cos\phi^0_{1/2}  +299\sin\phi^0_{1/2}) 
 -\frac{\langle\overline{q}q\rangle^2\langle\frac{\alpha_s}{\pi}G^2\rangle}{2^{9}3^3 \pi^2} (6\cos\phi^0_{1/2}  +61\sin\phi^0_{1/2}) \notag \\ 
&- \frac{13m_s\langle\frac{\alpha_s}{\pi}G^2\rangle\langle\overline{s}g\sigma \cdot G s\rangle}{2^{14}3 \pi^4}  \cos\phi^0_{1/2} 
- \frac{m_s\langle\overline{q}q\rangle^2\langle\overline{s}s\rangle}{2^33^2\pi^2} \sin\phi^0_{1/2}, \\
C_{12} =& \frac{2\langle\overline{q}q\rangle^4}{3^3}\sin\phi^0_{1/2} 
+\frac{\langle\overline{q}q\rangle\langle\overline{q}g\sigma \cdot G q\rangle\langle\frac{\alpha_s}{\pi}G^2\rangle}{2^{12}3^3\pi^2} (
65\cos\phi^0_{1/2} +418\sin\phi^0_{1/2}) \notag \\
&-\frac{m_s\langle\overline{q}q\rangle^2\langle\overline{s}g\sigma \cdot G s\rangle}{2^63^3\pi^2} (3\cos\phi^0_{1/2} +13\sin\phi^0_{1/2})
+\frac{7m_s\langle\overline{q}q\rangle \langle\overline{s}s\rangle\langle\overline{q}g\sigma \cdot G q\rangle}{2^53^2\pi^2} \sin\phi^0_{1/2},  \notag \\
C_{14}=&+\frac{31\langle\overline{q}q\rangle^3\langle\overline{q}g\sigma \cdot G q\rangle}{2^43^3}\sin\phi^0_{1/2} 
-\frac{m_s\langle\overline{q}q\rangle\langle\overline{q}g\sigma \cdot G q\rangle\langle\overline{s}g\sigma \cdot G s\rangle}{2^{9}3^3\pi^2} (
17\cos\phi^0_{1/2} +8\sin\phi^0_{1/2}) \notag\\
&+\frac{19m_s\langle\overline{s}s\rangle\langle\overline{q}g\sigma \cdot G q\rangle^2}{2^{9}3^2\pi^2}\sin\phi^0_{1/2} 
+\frac{19m_s\langle\overline{q}q\rangle^2\langle\overline{s}s\rangle\langle\frac{\alpha_s}{\pi}G^2\rangle}{2^73^4}\sin\phi^0_{1/2}. \notag
\end{align}

\subsubsection{Chiral-odd part}
\begin{align}
C_1 =& 0, \mspace{18mu} C_3 = \frac{\langle\overline{s}s\rangle}{2^{11}3^2 5\pi^6}\sin\phi^0_{1/2}, \mspace{18mu} 
C_5 = -\frac{7\langle\overline{s}g\sigma \cdot G s\rangle}{2^{14}3^2 \pi^6}\sin\phi^0_{1/2}, \mspace{18mu} 
C_7 = -\frac{5\langle\overline{s}s\rangle \langle\frac{\alpha_s}{\pi}G^2\rangle}{2^{11}3^3 \pi^4} \sin\phi^0_{1/2}, \notag\\
C_9 =& \frac{161\langle\frac{\alpha_s}{\pi}G^2\rangle\langle\overline{s}g\sigma \cdot G s\rangle}{2^{15}3^2 \pi^4} \sin\phi^0_{1/2}
 - \frac{\langle\overline{q}q\rangle^2\langle\overline{s}s\rangle}{2^23^2\pi^2}  \sin\phi^0_{1/2}, \notag \\ 
C_{11} =& \frac{7\langle\overline{q}q\rangle^2\langle\overline{s}g\sigma \cdot Gs\rangle}{2^63^2\pi^2} (\cos\phi^0_{1/2} +2\sin\phi^0_{1/2}) 
+ \frac{7\langle\overline{q}q\rangle \langle\overline{s}s\rangle\langle\overline{q}g\sigma \cdot Gq\rangle}{2^43^2\pi^2}\sin\phi^0_{1/2},  \\
C_{13} =& -\frac{25\langle\overline{q}q\rangle^2 \langle\overline{s}s\rangle \langle\frac{\alpha_s}{\pi}G^2\rangle}{2^63^3}\sin\phi^0_{1/2}  
-\frac{\langle\overline{q}q\rangle\langle\overline{q}g\sigma \cdot G q\rangle\langle\overline{s}g\sigma \cdot G s\rangle}{2^{12}3^3\pi^2}
(424\cos\phi^0_{1/2}+1979\sin\phi^0_{1/2}) \notag\\
& -\frac{19\langle\overline{s}s\rangle\langle\overline{q}g\sigma \cdot G q\rangle^2}{2^83^2\pi^2} \sin\phi^0_{1/2}. \notag
\end{align}

\subsection{$IJ^P = 1\frac{1}{2}^{\pm}$}
\subsubsection{Chiral-even part}
\begin{align}
C_0 =& 0, \mspace{18mu} C_4 = -\frac{ \langle\frac{\alpha_s}{\pi}G^2\rangle }{2^{14}3\cdot 5\pi^6}  \cos\phi^1_{1/2}, \mspace{18mu} 
C_6 = -\frac{\langle\overline{q}q\rangle^2}{2^63^2\pi^4}\sin\phi^1_{1/2} 
- \frac{m_s\langle\overline{s}g\sigma \cdot G s\rangle}{2^{13}\pi^6} \cos\phi^1_{1/2}, \notag \\
C_8 =& \frac{\langle\overline{q}q\rangle\langle\overline{q}g\sigma \cdot G q\rangle}{2^{9}3^2 \pi^4}  (3\cos\phi^1_{1/2} +44\sin\phi^1_{1/2}), \notag\\
C_{10} =& -\frac{\langle\overline{q}g\sigma \cdot G q\rangle^2}{2^{13}3^2 \pi^4} (15\cos\phi^1_{1/2}  +361\sin\phi^1_{1/2}) 
+\frac{\langle\overline{q}q\rangle^2\langle\frac{\alpha_s}{\pi}G^2\rangle}{2^{9}3^3 \pi^2} (18\cos\phi^1_{1/2}  -29\sin\phi^1_{1/2}) \notag \\ 
&- \frac{13m_s\langle\frac{\alpha_s}{\pi}G^2\rangle\langle\overline{s}g\sigma \cdot G s\rangle}{2^{14}3 \pi^4}  \cos\phi^1_{1/2} 
- \frac{m_s\langle\overline{q}q\rangle^2\langle\overline{s}s\rangle}{2^23^2\pi^2} \sin\phi^1_{1/2}, \\
C_{12} =& -\frac{2\langle\overline{q}q\rangle^4}{3^3}\sin\phi^1_{1/2} 
-\frac{\langle\overline{q}q\rangle\langle\overline{q}g\sigma \cdot G q\rangle\langle\frac{\alpha_s}{\pi}G^2\rangle}{2^{12}3^3\pi^2} (
135\cos\phi^1_{1/2} -158\sin\phi^1_{1/2}) \notag \displaybreak[4] \\
&+\frac{m_s\langle\overline{q}q\rangle^2\langle\overline{s}g\sigma \cdot G s\rangle}{2^63^3\pi^2} (9\cos\phi^1_{1/2} +13\sin\phi^1_{1/2}) 
+\frac{19m_s\langle\overline{q}q\rangle \langle\overline{s}s\rangle\langle\overline{q}g\sigma \cdot G q\rangle}{2^63^2\pi^2} \sin\phi^1_{1/2},  \notag \\
C_{14}=&-\frac{31\langle\overline{q}q\rangle^3\langle\overline{q}g\sigma \cdot G q\rangle}{2^43^3}\sin\phi^1_{1/2} 
+\frac{m_s\langle\overline{q}q\rangle\langle\overline{q}g\sigma \cdot G q\rangle\langle\overline{s}g\sigma \cdot G s\rangle}{2^{10}3^3\pi^2} (
69\cos\phi^1_{1/2} +122\sin\phi^1_{1/2}) \notag\\
&+\frac{11m_s\langle\overline{s}s\rangle\langle\overline{q}g\sigma \cdot G q\rangle^2}{2^{8}3^2\pi^2}\sin\phi^1_{1/2} 
+\frac{23m_s\langle\overline{q}q\rangle^2\langle\overline{s}s\rangle\langle\frac{\alpha_s}{\pi}G^2\rangle}{2^73^4}\sin\phi^1_{1/2}. \notag
\end{align}

\subsubsection{Chiral-odd part}
\begin{align}
C_1 =& 0, \mspace{18mu} C_3 = -\frac{\langle\overline{s}s\rangle}{2^{11}3^2 5\pi^6}\sin\phi^1_{1/2}, \mspace{18mu} 
C_5 = +\frac{7\langle\overline{s}g\sigma \cdot G s\rangle}{2^{14}3^2 \pi^6}\sin\phi^1_{1/2}, \mspace{18mu} 
C_7 = +\frac{5\langle\overline{s}s\rangle \langle\frac{\alpha_s}{\pi}G^2\rangle}{2^{11}3^3 \pi^4} \sin\phi^1_{1/2}, \notag \\
C_9 =& -\frac{161\langle\frac{\alpha_s}{\pi}G^2\rangle\langle\overline{s}g\sigma \cdot G s\rangle}{2^{15}3^2 \pi^4} \sin\phi^1_{1/2}
 - \frac{\langle\overline{q}q\rangle^2\langle\overline{s}s\rangle}{2 \cdot 3^2\pi^2}  \sin\phi^1_{1/2},  \\ 
C_{11} =& \frac{\langle\overline{q}q\rangle^2\langle\overline{s}g\sigma \cdot Gs\rangle}{2^63^2\pi^2} (3\cos\phi^1_{1/2} +19\sin\phi^1_{1/2}) 
+ \frac{19\langle\overline{q}q\rangle \langle\overline{s}s\rangle\langle\overline{q}g\sigma \cdot Gq\rangle}{2^53^2\pi^2}\sin\phi^1_{1/2}, \notag \\
C_{13} =& -\frac{29\langle\overline{q}q\rangle^2 \langle\overline{s}s\rangle \langle\frac{\alpha_s}{\pi}G^2\rangle}{2^63^3}\sin\phi^1_{1/2} 
-\frac{\langle\overline{q}q\rangle\langle\overline{q}g\sigma \cdot G q\rangle\langle\overline{s}g\sigma \cdot G s\rangle}{2^{12}3^3\pi^2}
(204\cos\phi^1_{1/2}+2503\sin\phi^1_{1/2}) \notag\\
& -\frac{11\langle\overline{s}s\rangle\langle\overline{q}g\sigma \cdot G q\rangle^2}{2^73^2\pi^2} \sin\phi^1_{1/2}. \notag
\end{align}

\subsection{$IJ^P = 0\frac{3}{2}^{\pm}$}
\begin{center}
See \cite{Gubler}.
\end{center}

\subsection{$IJ^P = 1\frac{3}{2}^{\pm}$}
\subsubsection{Chiral-even part}
\begin{align}
C_0 =& 0, \mspace{18mu} C_4 = \frac{\langle\frac{\alpha_s}{\pi}G^2\rangle }{2^{16}3^3\cdot 5\pi^6}  \cos\phi^1_{3/2}, \mspace{18mu} 
C_6 = \frac{\langle\overline{q}q\rangle^2}{2^63^2 5\pi^4}\sin\phi^1_{3/2} 
+ \frac{m_s\langle\overline{s}g\sigma \cdot G s\rangle}{2^{14}3 \cdot 5\pi^6} \cos\phi^1_{3/2}, \notag \\
C_8 =& -\frac{\langle\overline{q}q\rangle\langle\overline{q}g\sigma \cdot G q\rangle}{2^{12}3^2 \pi^4}  (\cos\phi^1_{3/2} +54\sin\phi^1_{3/2}), \notag\\
C_{10} =& \frac{\langle\overline{q}g\sigma \cdot G q\rangle^2}{2^{14}3^4 \pi^4} (15\cos\phi^1_{3/2}  +677\sin\phi^1_{3/2}) 
 -\frac{\langle\overline{q}q\rangle^2\langle\frac{\alpha_s}{\pi}G^2\rangle}{2^{10}3^3 \pi^2} (2\cos\phi^1_{3/2}  +11\sin\phi^1_{3/2}) \notag\\
&+ \frac{13m_s\langle\frac{\alpha_s}{\pi}G^2\rangle\langle\overline{s}g\sigma \cdot G s\rangle}{2^{15}3^3 \pi^4}  \cos\phi^1_{3/2} 
+ \frac{m_s\langle\overline{q}q\rangle^2\langle\overline{s}s\rangle}{2^33^3\pi^2} \sin\phi^1_{3/2}, \\
C_{12} =& \frac{\langle\overline{q}q\rangle^4}{3^3}\sin\phi^1_{3/2} 
+\frac{5\langle\overline{q}q\rangle\langle\overline{q}g\sigma \cdot G q\rangle\langle\frac{\alpha_s}{\pi}G^2\rangle}{2^{14}3^2\pi^2} (
3\cos\phi^1_{3/2} +20\sin\phi^1_{3/2}) \notag\\
&-\frac{m_s\langle\overline{q}q\rangle^2\langle\overline{s}g\sigma \cdot G s\rangle}{2^83^3\pi^2} (3\cos\phi^1_{3/2} +10\sin\phi^1_{3/2}) 
-\frac{11m_s\langle\overline{q}q\rangle \langle\overline{s}s\rangle\langle\overline{q}g\sigma \cdot G q\rangle}{2^83^2\pi^2} \sin\phi^1_{3/2},  \notag\\
C_{14}=& \frac{97\langle\overline{q}q\rangle^3\langle\overline{q}g\sigma \cdot G q\rangle}{2^53^4}\sin\phi^1_{3/2} 
-\frac{m_s\langle\overline{q}q\rangle\langle\overline{q}g\sigma \cdot G q\rangle\langle\overline{s}g\sigma \cdot G s\rangle}{2^{11}3^3\pi^2} (
23\cos\phi^1_{3/2} +18\sin\phi^1_{3/2}) \notag\\
&-\frac{11m_s\langle\overline{s}s\rangle\langle\overline{q}g\sigma \cdot G q\rangle^2}{2^{9}3^4\pi^2}\sin\phi^1_{3/2} 
+\frac{25m_s\langle\overline{q}q\rangle^2\langle\overline{s}s\rangle\langle\frac{\alpha_s}{\pi}G^2\rangle}{2^83^4}\sin\phi^1_{3/2}. \notag
\end{align}

\subsubsection{Chiral-odd part}
\begin{align}
C_1 =& 0, \mspace{18mu} C_3 = -\frac{\langle\overline{s}s\rangle}{2^{13}3^3 \pi^6}\sin\phi^1_{3/2}, \mspace{18mu} 
C_5 = +\frac{5\langle\overline{s}g\sigma \cdot G s\rangle}{2^{13}3^3 \pi^6}\sin\phi^1_{3/2}, \mspace{18mu} 
C_7 = -\frac{7\langle\overline{s}s\rangle \langle\frac{\alpha_s}{\pi}G^2\rangle}{2^{14}3^3 \pi^4} \sin\phi^1_{3/2}, \notag \\
C_9 =& +\frac{5\langle\frac{\alpha_s}{\pi}G^2\rangle\langle\overline{s}g\sigma \cdot G s\rangle}{2^{15}3^4 \pi^4} \sin\phi^1_{3/2}
 - \frac{\langle\overline{q}q\rangle^2\langle\overline{s}s\rangle}{2^3 3^2\pi^2}  \sin\phi^1_{3/2}, \\ 
C_{11} =& -\frac{\langle\overline{q}q\rangle^2\langle\overline{s}g\sigma \cdot Gs\rangle}{2^83^2\pi^2} (\cos\phi^1_{3/2} -15\sin\phi^1_{3/2}) \label{eq:odd4}  
+ \frac{19\langle\overline{q}q\rangle \langle\overline{s}s\rangle\langle\overline{q}g\sigma \cdot Gq\rangle}{2^73^2\pi^2}\sin\phi^1_{3/2}, \notag \\
C_{13} =& -\frac{7\langle\overline{q}q\rangle^2 \langle\overline{s}s\rangle \langle\frac{\alpha_s}{\pi}G^2\rangle}{2^63^3}\sin\phi^1_{3/2} 
+\frac{\langle\overline{q}q\rangle\langle\overline{q}g\sigma \cdot G q\rangle\langle\overline{s}g\sigma \cdot G s\rangle}{2^{11}3^3\pi^2}
(17\cos\phi^1_{3/2}-216\sin\phi^1_{3/2}) \notag\\
&-\frac{11\langle\overline{s}s\rangle\langle\overline{q}g\sigma \cdot G q\rangle^2}{2^63^4\pi^2} \sin\phi^1_{3/2}. \notag
\end{align}

\section{Establishment of a valid Borel window \label{apB}}
In this Appendix, we explicitly show that a Borel window has been obtained for the sum rules of the chiral-even part for the various 
quantum numbers. 

\begin{figure*}
\begin{center}
\includegraphics[width=8cm,clip]{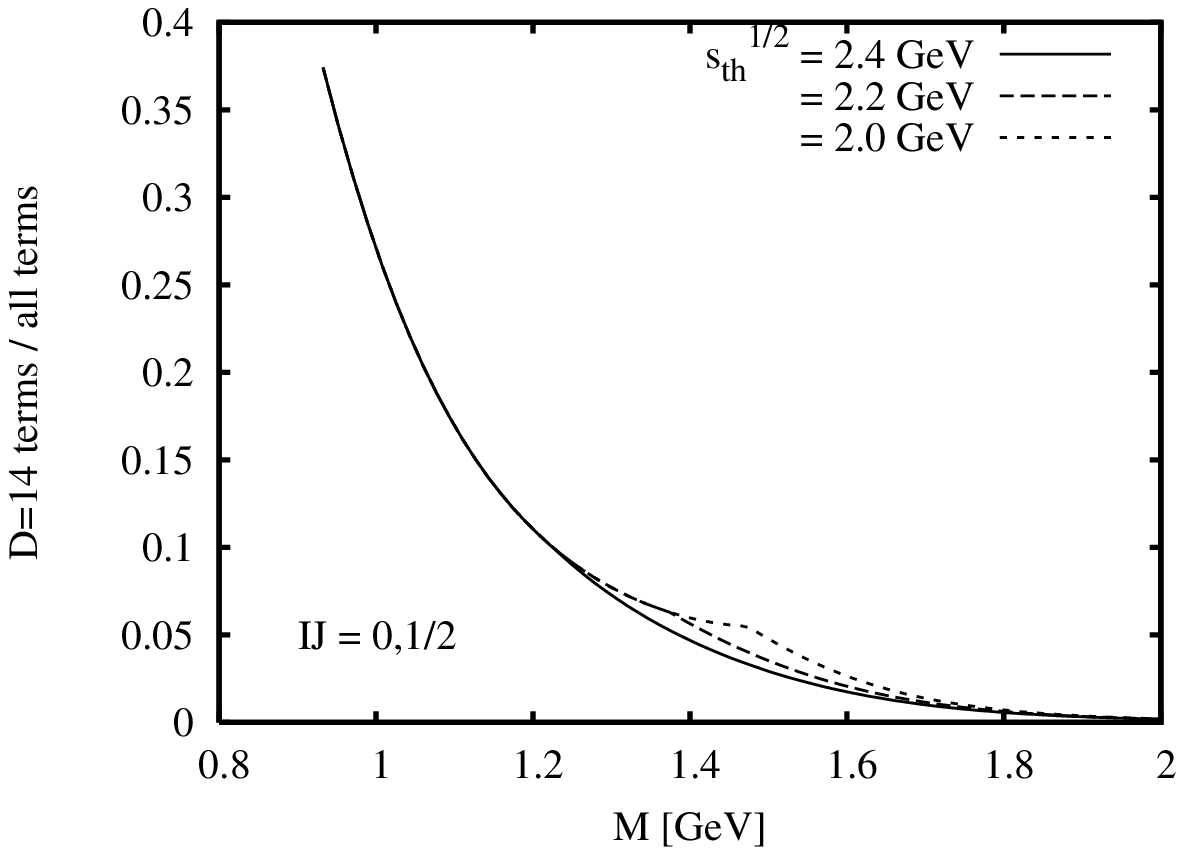}
\includegraphics[width=8cm,clip]{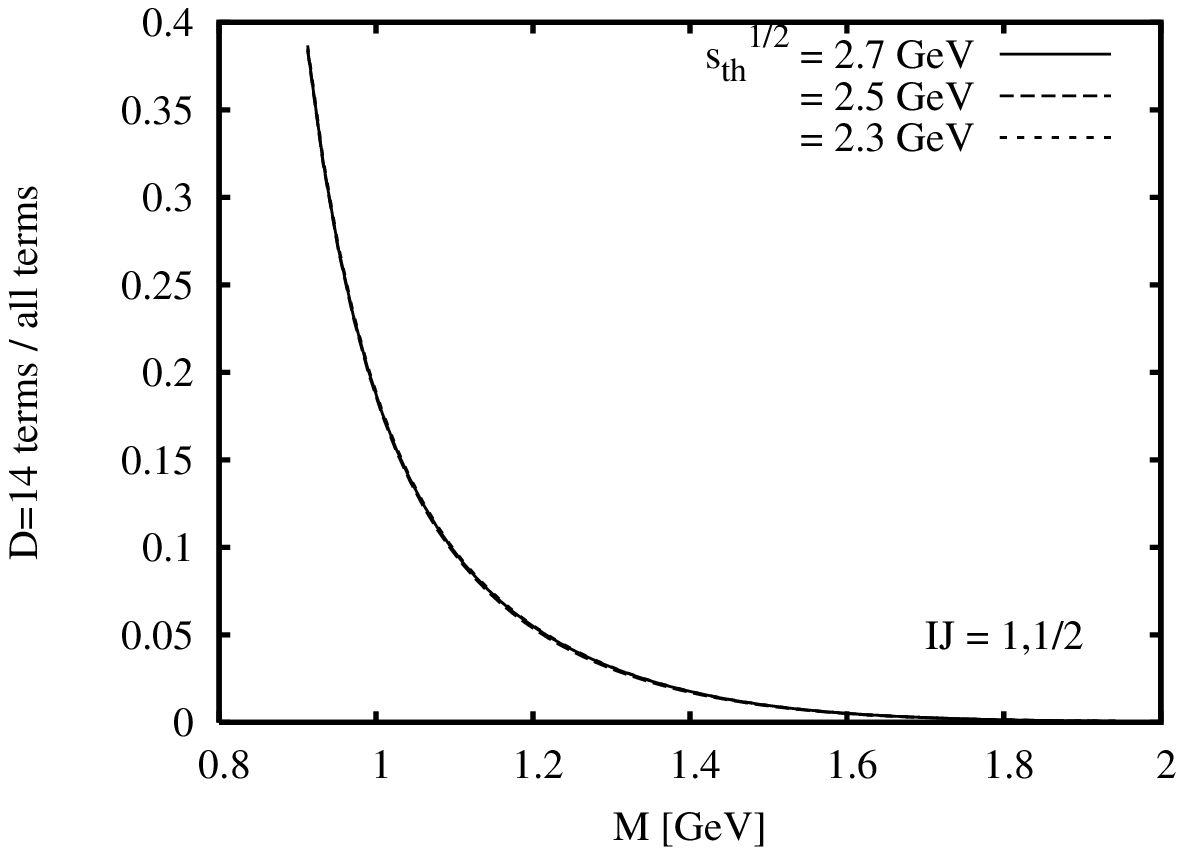}
\includegraphics[width=8cm,clip]{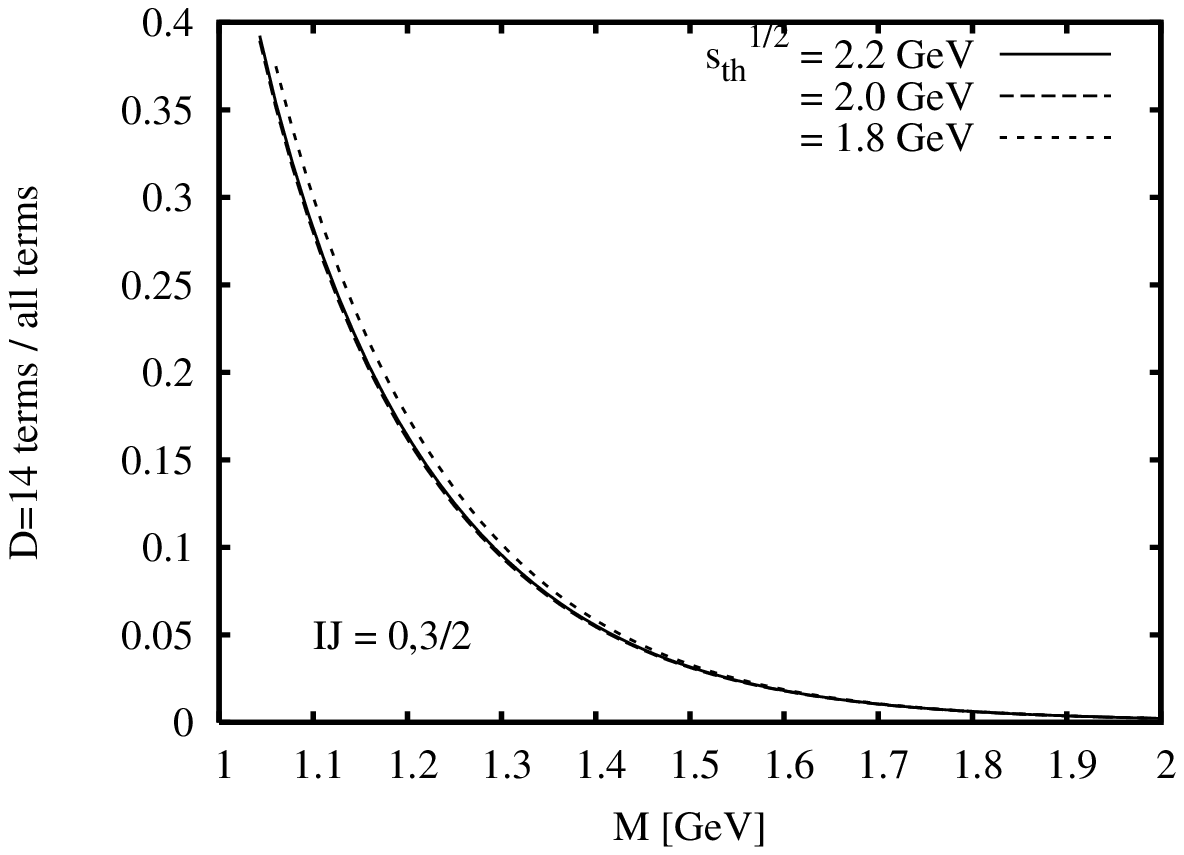}
\includegraphics[width=8cm,clip]{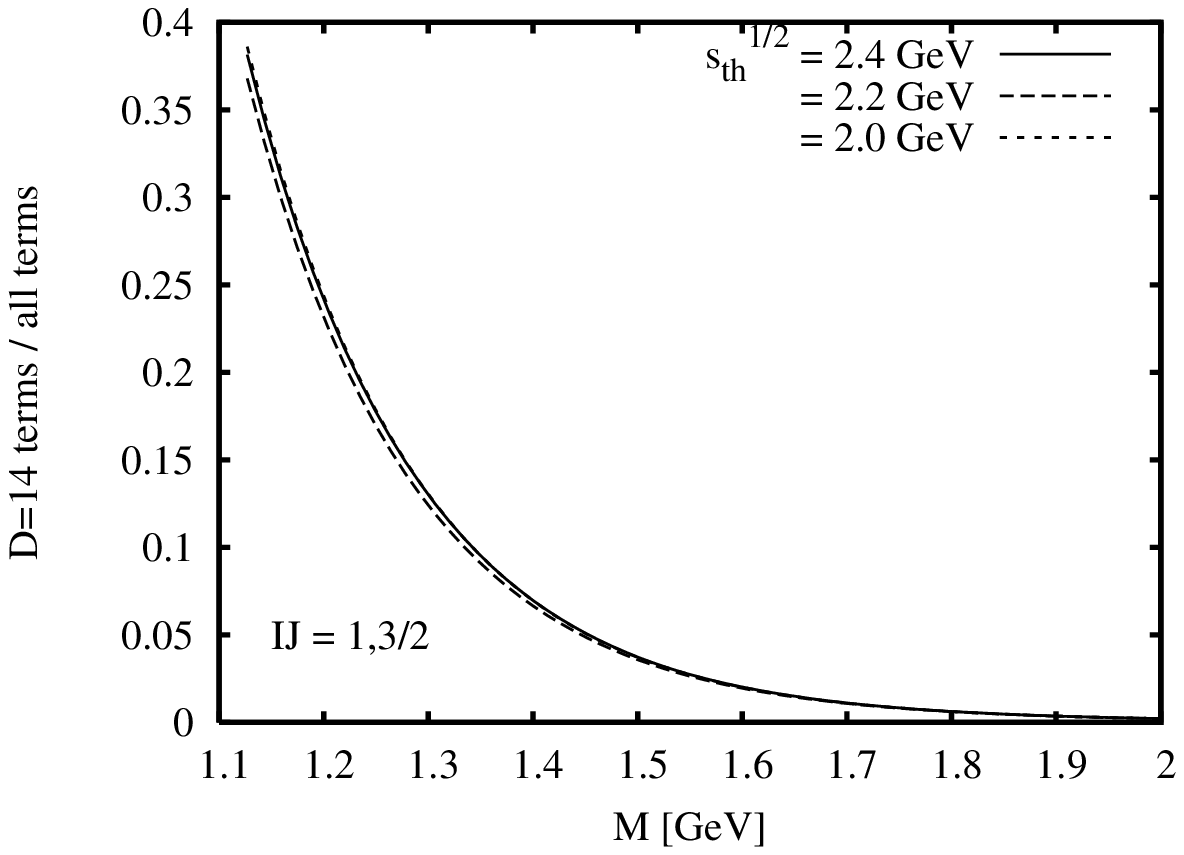}
\caption{The highest term in the OPE divided by the whole OPE for $IJ = 0\frac{1}{2}$, 
$IJ = 1\frac{1}{2}$, $IJ = 0\frac{3}{2}$, $IJ = 1\frac{3}{2}$, 
as given in the left-hand side of Eq.(\ref{eq:conv}).}
\label{fig:conv1}
\end{center}
\end{figure*}

\begin{figure*}
\begin{center}
\includegraphics[width=8cm,clip]{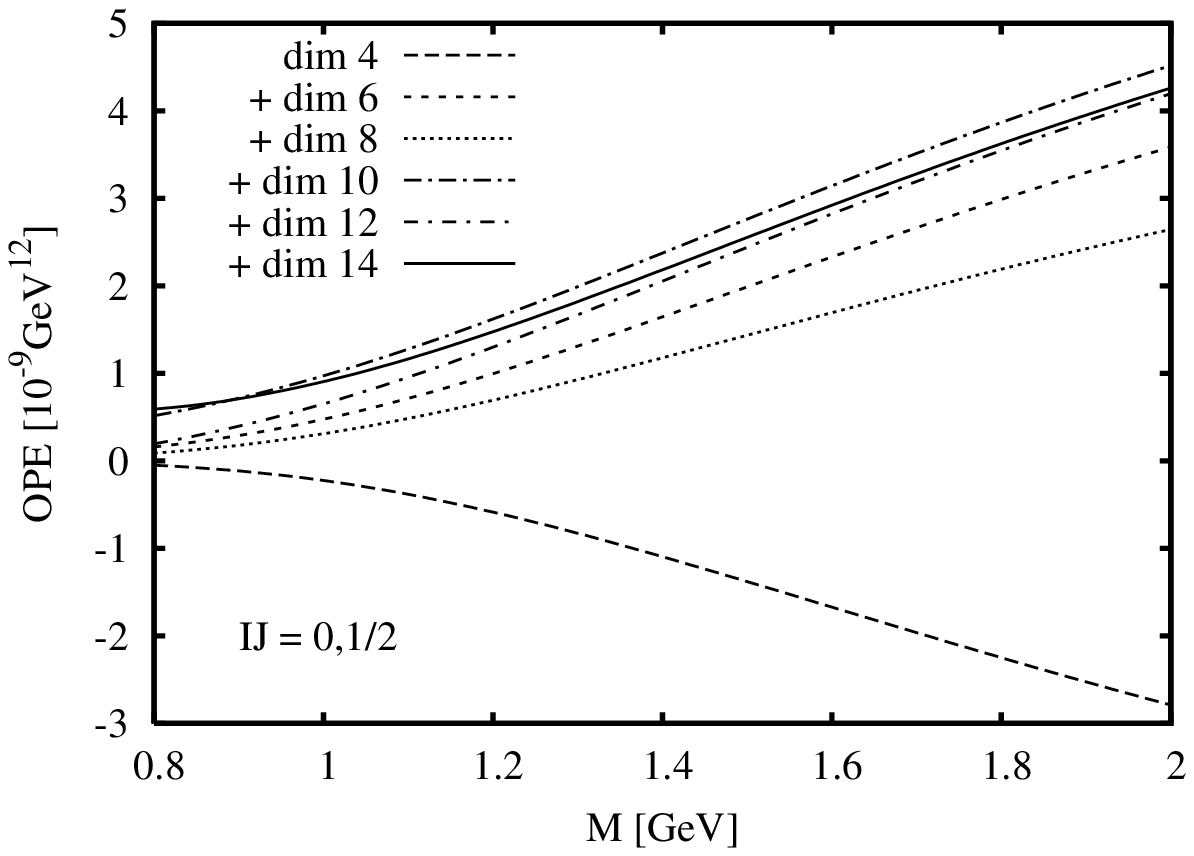}
\includegraphics[width=8cm,clip]{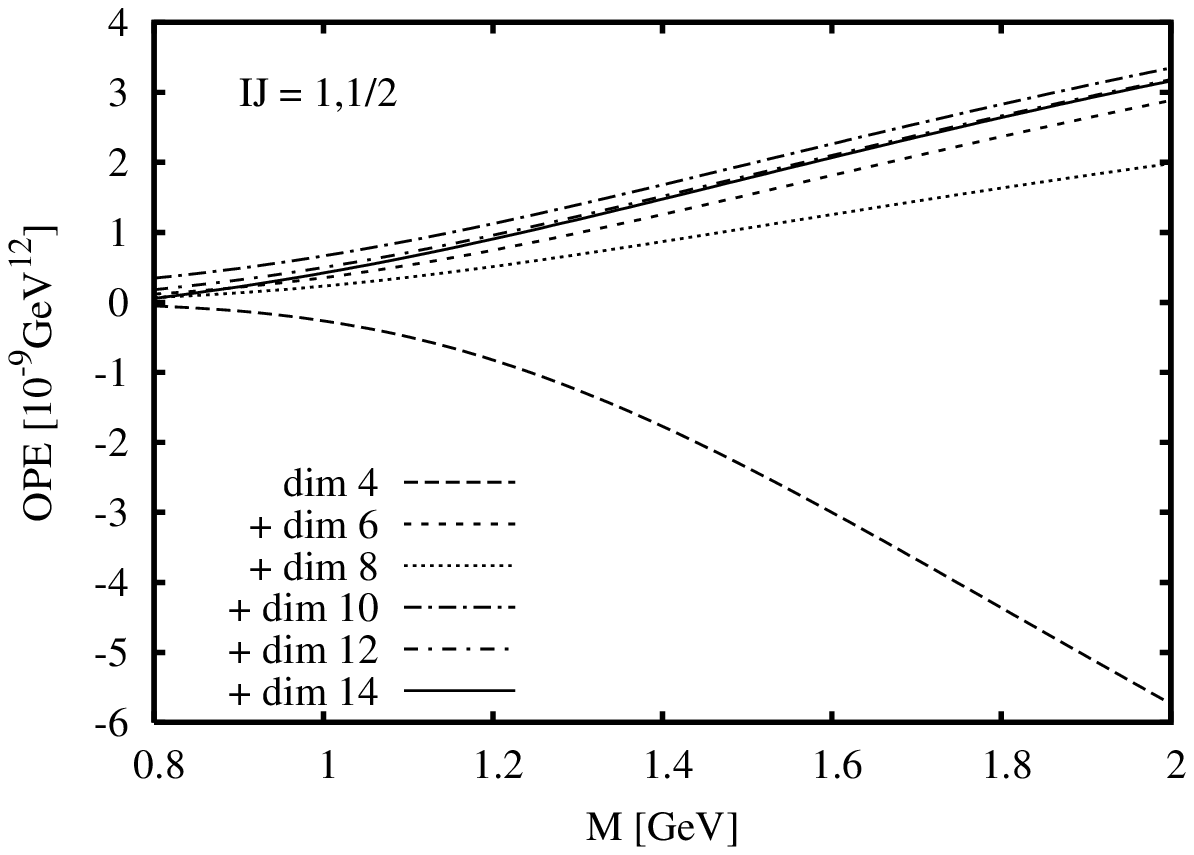}
\includegraphics[width=8cm,clip]{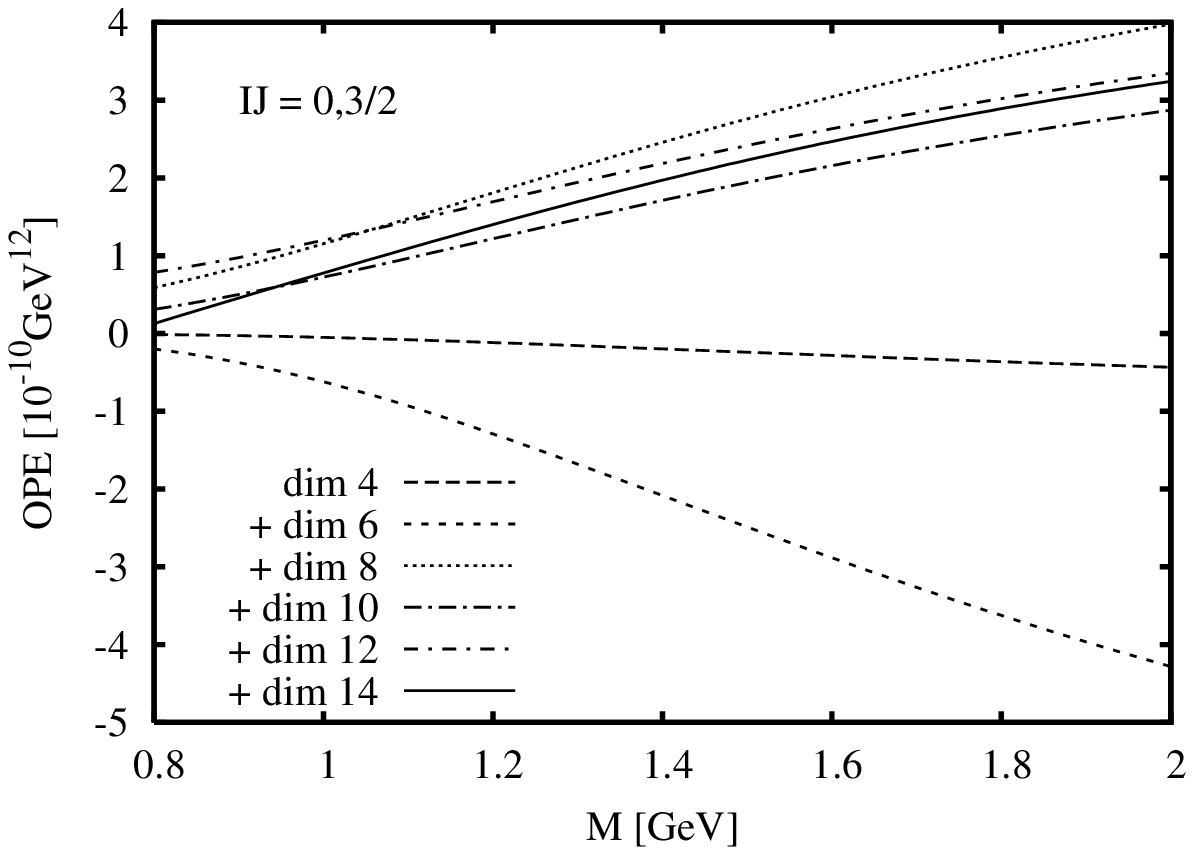}
\includegraphics[width=8cm,clip]{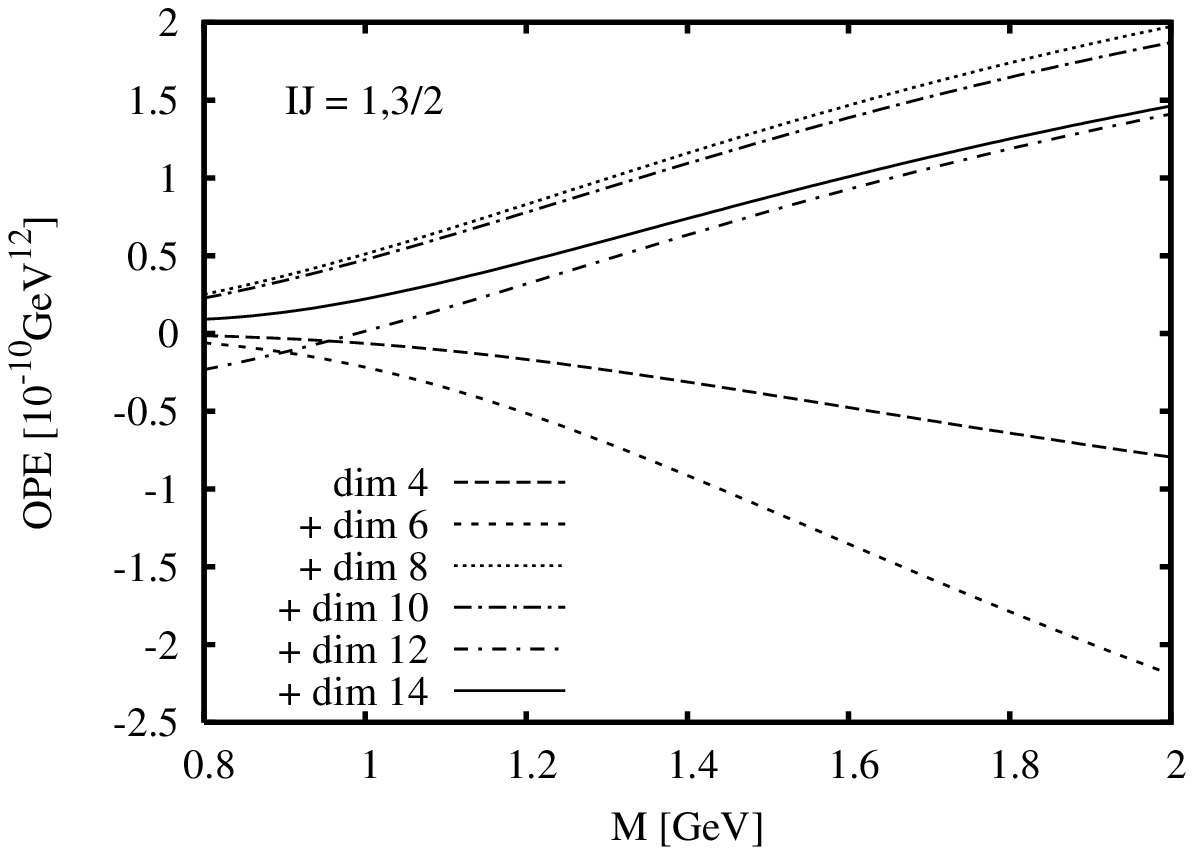}
\caption{Contributions of different dimensions to the right-hand side of Eq.(\ref{eq:ope1}) for $IJ = 0\frac{1}{2}$, 
$IJ = 1\frac{1}{2}$, $IJ = 0\frac{3}{2}$, $IJ = 1\frac{3}{2}$, added in succession. 
The expansion in most cases starts to converge after terms up to dimension 10 are included.}
\label{fig:ope1}
\end{center}
\end{figure*}

\begin{figure*}
\begin{center}
\includegraphics[width=8cm,clip]{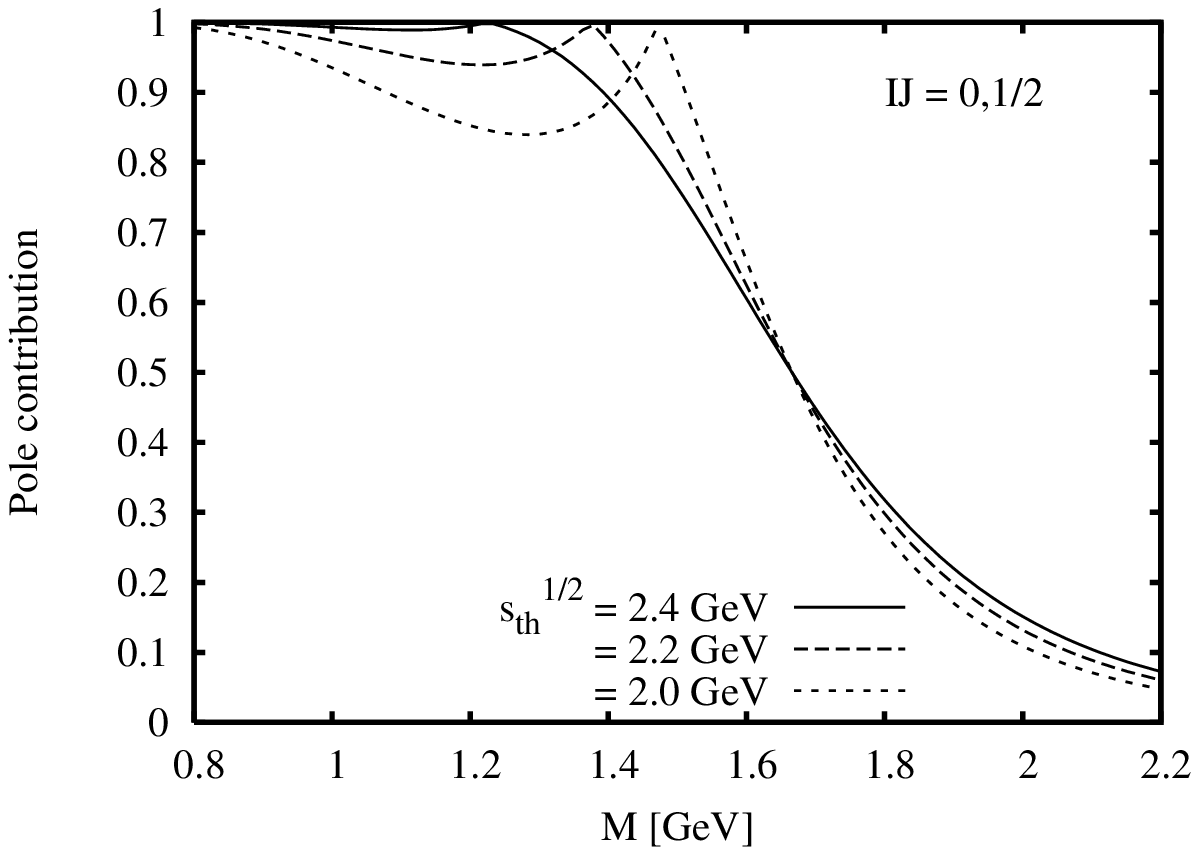}
\includegraphics[width=8cm,clip]{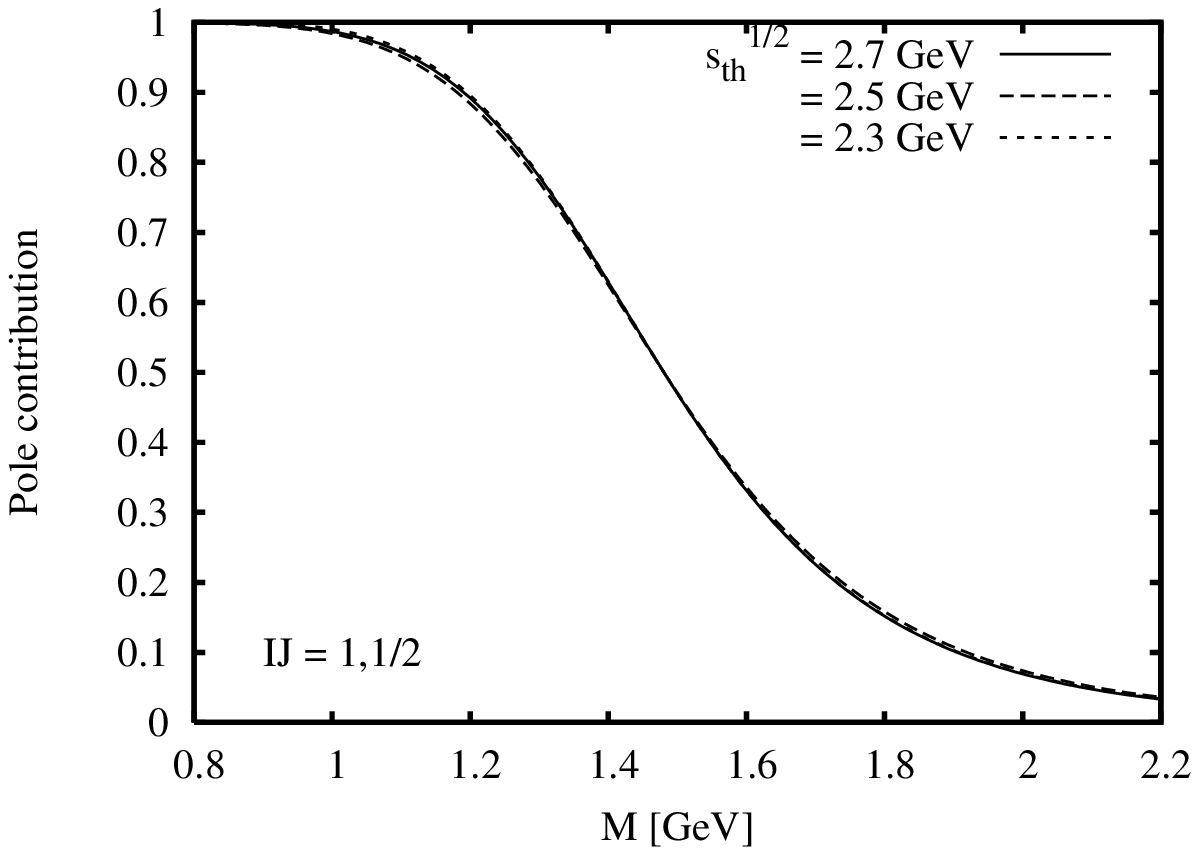}
\includegraphics[width=8cm,clip]{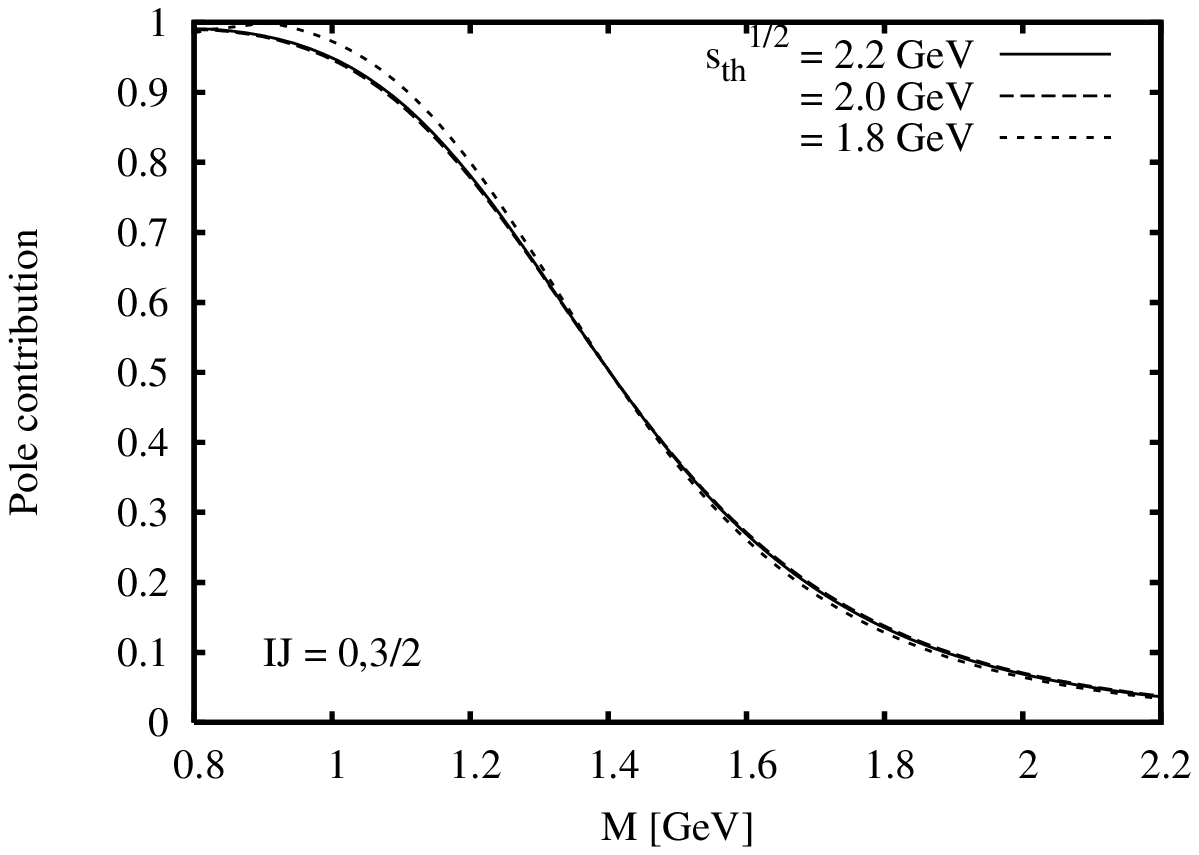}
\includegraphics[width=8cm,clip]{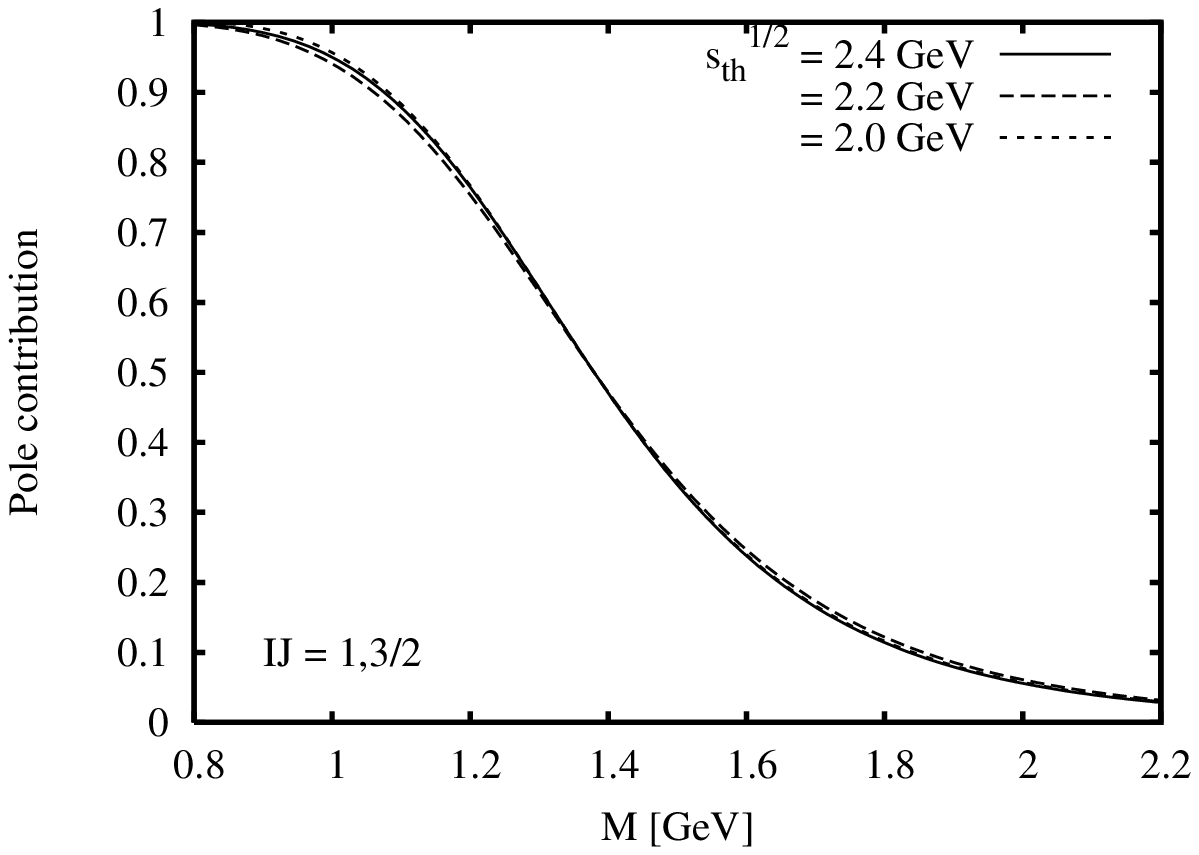}
\caption{The pole contribution to the sum rule with $IJ = 0\frac{1}{2}$ 
$IJ = 1\frac{1}{2}$, $IJ = 0\frac{3}{2}$, $IJ = 1\frac{3}{2}$ 
in comparison with the contribution 
of the high-energy continuum states. 
These plots, together with the ones in Fig. \ref{fig:conv1} finally lead to the Borel windows 
referred to in the result section.}
\label{fig:pole1}
\end{center}
\end{figure*}

First, the convergence of the OPE is checked. 
This is done by calculating the left-hand side of Eq.(\ref{eq:conv}). The results are given in Fig. \ref{fig:conv1}. Additionally, 
the right-hand side of Eq.(\ref{eq:ope1}) added order by order is shown in Fig. \ref{fig:ope1} 
to get a better idea of the behavior of the expansion. 
Subsequently, the pole contribution of Eq.(\ref{eq:polecontr}) is investigated. The corresponding plots are given in 
Fig. \ref{fig:pole1} for the various quantum numbers. 

\twocolumngrid


\end{document}